\documentclass{article}

\usepackage{graphicx} 
\usepackage[a4paper,top=25mm,bottom=25mm]{geometry}
\usepackage[font=footnotesize, labelfont=bf]{caption}
\usepackage{subcaption} 
\usepackage{setspace}
\usepackage{amssymb}
\usepackage{amsmath}
\usepackage{verbatim} 
\usepackage{csquotes} 
\usepackage{float} 
\usepackage{dcolumn}
\usepackage{authblk}
\usepackage[hypertexnames=false]{hyperref}
\usepackage[table]{xcolor}
\usepackage{tabularx}
\usepackage{svg} 
\usepackage{dsfont} 

\usepackage{array}
\usepackage{longtable} 

\usepackage{booktabs}
\usepackage{makecell}
\usepackage{threeparttable}
\usepackage{multirow}
\usepackage{rotating}
\usepackage{adjustbox}

\usepackage[
backend=biber,
style=science,
sorting=none
]{biblatex}
\definecolor{clustergray}{gray}{0.9}
\definecolor{stringwhite}{rgb}{1,1,1}

\title{The Costs of Early-career Disciplinary Pivots: \\Evidence from PhD Admissions}

\author[1,$\dagger$]{Sidney Xiang}
\author[2,$\dagger$]{Nicholas David}
\author[1]{Dallas Card}
\author[2]{Wenhao Sun}
\author[1]{Daniel M Romero}
\author[1,*]{Misha Teplitskiy}

\affil[1]{School of Information, University of Michigan}
\affil[2]{Department of Materials Science and Engineering, University of Michigan}
\affil[$\dagger$]{These authors contributed equally to this work.}
\affil[*]{Corresponding author: \href{mailto:tepl@umich.edu}{tepl@umich.edu}}

\date{\today}

\makeatletter
\let\SI@maketitle\maketitle
\let\SI@@maketitle\@maketitle
\let\SI@and\and
\let\SI@thanks\thanks
\makeatother

\begin{document}

\makeatletter
\global\let\SI@author\@author
\global\let\SI@date\@date
\makeatother

\maketitle

\begin{abstract}
Science often advances when researchers pivot from one discipline to another and import new ideas. However, for established researchers, such pivots are associated with fewer publications and citations. Here, we investigate pivoting at the beginning of a research career—doctoral admissions—when the benefits of importing ideas might outweigh the switching costs.
Analyzing 171,347 PhD applications at a large research university from 2013 to 2023, we measured the semantic distance between each applicant's prior degree and target program. Applicants who pivot further had lower undergraduate grades and, controlling for academic and demographic predictors, lower admission chances: the average pivot corresponds to a 5.3\% lower admission probability (0.80 percentage points). Among enrollees, the average pivoter takes longer to graduate than non-pivoters, with a 10.7\% lower graduation hazard, and similar publication and citation counts on average and at the tails. Our results suggest that pivoting at the outset of research careers, even if it benefits research communities, comes with significant costs to individuals.
\end{abstract}

\section{Introduction}

Research communities are often enriched by outsiders. For example, physicists who turned to biology brought quantitative and structural approaches that helped launch molecular biology \cite{morangeHistoryMolecularBiology1998}; psychologists who entered economics reshaped the discipline's understanding of decision-making \cite{camererBehavioralEconomicsPast2004}; and computer scientists who moved into the social sciences introduced computational frameworks that advanced theory \cite{edelmannComputationalSocialScience2020}. More generally, by introducing unfamiliar methods, data, and theories, disciplinary pivoters can help a community combine these knowledge elements with more familiar ones to expand its solution space to existing problems \cite{jeppesen2010marginality, hongGroupsDiverseProblem2004} and create more impactful contributions \cite{shiSurprisingCombinationsResearch2023, uzziAtypicalCombinationsScientific2013}.

Despite the potential for community benefits, pivoters can face significant barriers individually. Recent work by Hill et al. \cite{hillPivotPenaltyResearch2025} documents lower publication success and citation impact for researchers pivoting from one research community to another. They find support for the penalty being explained by the entrenchment of expertise, reputation, and professional networks in the old community, a finding that is echoed in the literature on interdisciplinary research evaluation \cite{leaheyPerksPerilsInterdisciplinary2018,finiNewTakeCategorical2023,xiang2025evaluating} and the evaluation of category-spanning producers more generally \cite{hsuJacksAllTrades2006, kovacsStickinessCategoryLabels2024, leungDilettanteRenaissancePerson2014, zuckermanRobustIdentitiesNonentities2003}. Scientists' observed choices are consistent with these significant penalties: they tend to avoid large pivots \cite{jiaQuantifyingPatternsResearchinterest2017}, even when incentivized by substantial topic-specific grant funding \cite{myersElasticityScience2020}. This prior work raises a natural question: when, if ever, is disciplinary pivoting rewarded?

Doctoral admissions is a compelling setting to examine. First, we show that pivoting at this stage is very common, with 38.5\% of applicants targeting programs outside of their prior disciplinary training. This prevalence motivates two competing predictions. One possibility is that pivoters benefit in admissions: committees value fresh perspectives and cognitive diversity, and applicants not yet deeply specialized can bring approaches from their home field to the target one. If so, we would expect pivoters to face no admission penalty, or even a premium, and to outperform non-pivoters after enrollment. The opposing possibility is that applicants accumulate field-specific knowledge through prior coursework and training that is unlikely to fully transfer across disciplines \cite{beckerInvestmentHumanCapital1962,jonesBurdenKnowledgeDeath2009a}, putting pivoters at a disadvantage in meeting program requirements on time. Faculty may weigh attrition risk heavily and pass over high-quality but unconventional candidates in favor of those perceived as safer bets \cite{posselt2016inside}. Our first research question is: Does there exist a pivot penalty in doctoral admissions?

Even if there is a pivot penalty in admissions, whether pivoters who do get admitted and enroll outperform non-pivoters is a priori ambiguous. On one hand, pivoters may show higher-variance outcomes---lower average performance but higher upper-tail impact---consistent with boundary-crossing research strategies in biomedical research \cite{foster2015tradition}, interdisciplinary research \cite{leaheyNotProductivityAlone2007} and patenting \cite{fleming2001recombinant}. On the other hand, pivoters may also underperform after enrollment, since the same deficits in field-specific training that may impede admission also impede degree completion and early-career research productivity. This motivates our second research question: Do pivoters, once enrolled, have different academic performance than non-pivoters?

We answer these questions using a dataset of all 171,347 PhD applications to a large public research university between 2013 and 2023. We operationalize pivots as applications where the applicant's prior degree does not match the target doctoral degree. Because some pivots span a greater disciplinary distance than others---moving from chemistry to biochemistry is not the same as moving from history to physics---we further characterize each application with a continuous measure of pivot distance that captures how far intellectually the applicant's prior field lies from the target doctoral field. This is a non-trivial task; identifying a pivot requires resolving different degree programs across thousands of different institutions to University X's doctoral degree programs. Furthermore, prior degree programs that appear similar by name may prepare students differently with curricula that vary in how well they train applicants for research. To resolve these concerns, we develop three complementary operationalizations of the pivot---the mentioned intellectual pivot distance, a behavioral measure which tracks observed field transitions, and a simpler binary indicator of field mismatch.  

Across all three pivot measures we find that, conditional on academic achievement, program-year selectivity, and demographic characteristics, greater pivot distance is associated with lower admission probability---the average pivot corresponds to a 5.3\% relative decline in the probability of admission (0.80 percentage points against a 15.0\% baseline acceptance rate). Among enrollees, the same contrast predicts a 10.7\% lower graduation hazard, with no evidence of a publication or citation advantage to offset it, whether on average or at the upper tail.

Our results make three contributions. First, we extend the literature on pivot penalties in science \cite{hillPivotPenaltyResearch2025} by showing that such penalties emerge at the earliest stage of the academic pipeline---doctoral admissions---before applicants have developed the expertise, professional networks, and reputations whose loss is thought to drive pivot penalties among established scientists. Pivot penalties are thus not solely a consequence of career-long switching costs, but also reflect more fundamental challenges of crossing disciplinary boundaries, such as a lack of field-specific training or perceived fit. Second, we present suggestive evidence that a recommendation letter writer who publishes in the target discipline attenuates the admissions pivot penalty by roughly three quarters. Even this early, where reputation and networks are thin, signals of connection to the target community appear to help overcome evaluators' concerns, linking the pivot penalty to broader work on bridging ties in knowledge transfer \cite{artsParadiseNoveltyLoss2018}. Third, we find that early-career pivoting does not reliably confer a compensating upper-tail advantage: pivoters who enroll are less likely to graduate and no more likely to achieve exceptional publication or citation impact. Taken together, disciplinary boundaries constrain the flow of new ideas into research communities not only through penalties on established scientists, but also through barriers at the earliest point of entry into a field.

\section{Results}

\subsection{Descriptives and adverse selection}

In our data, 38.5\% of doctoral applicants applied to a program in a different subject area than their most recent prior degree (i.e., pivoted). However, some of these pivots are greater than others (e.g., Biology $\rightarrow$ Cellular/molecular biology vs. Physics $\rightarrow$ Sociology). We measure the size of each pivot with an intellectual pivot distance $d$, a text-based dissimilarity between the Web of Science subject categories of an applicant's prior degree and their applied PhD program, taken from published estimates derived from the text of roughly $10^{7}$ publication abstracts \cite{diasUsingText2018} (Materials and Methods, \ref{sec:pivot}; SI Section~\ref{sec:intellectual}). For descriptive purposes, we group applications into non-pivoters ($d = 0$; 61.5\% of applications), smaller pivoters ($0 < d \le \Delta$; 20.6\%), and larger pivoters ($d > \Delta$; 17.1\%), where $\Delta = 0.305$ is the average-pivot contrast defined as the difference in mean pivot distance between pivoters and non-pivoters; because non-pivoters take $d = 0$ under this measure, it is simply the mean distance among pivoters---the ``average'' pivot. The remaining 0.9\% (1,610 applications) are pivots whose distance cannot be measured, because the prior or target specialty is one of the 29 for which no published distance exists (SI Section~\ref{sec:measures})

\begin{figure}[!htb]
    \centering
        \includegraphics[width=0.8\linewidth]{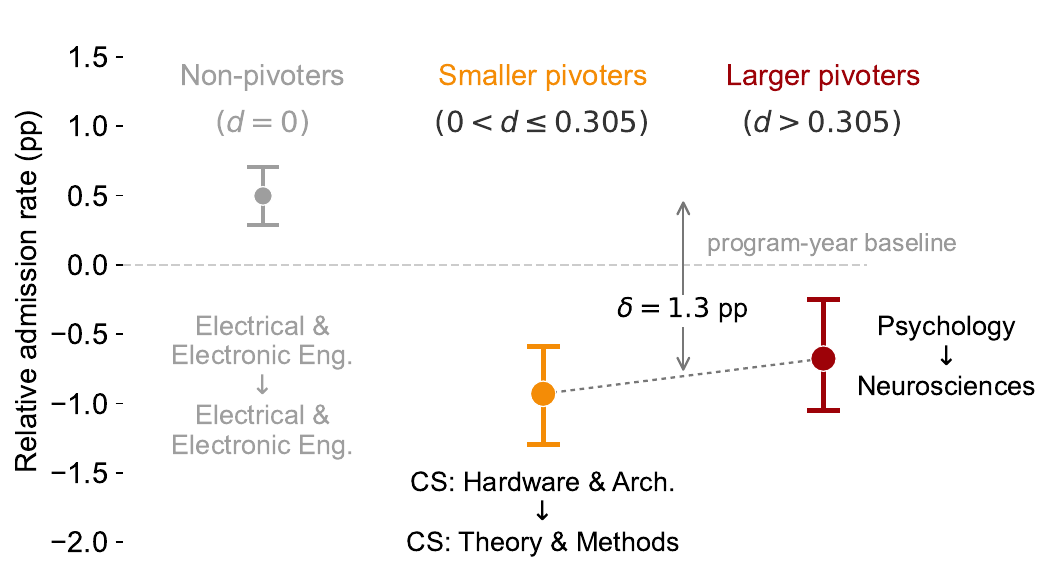}
    \caption{\textbf{Pivoters are admitted at lower rates than their non-pivoting peers.} Admission is expressed relative to the program-year baseline: each application's admission outcome is centered on the mean admission rate of the program and cycle to which it was submitted, so a value of zero (dashed line) denotes the average applicant to that program-year and each group is compared net of which programs it applied to. Applications are grouped by intellectual pivot distance $d$ into non-pivoters ($d = 0$), smaller pivoters ($0 < d \le \Delta$), and larger pivoters ($d > \Delta$), where $\Delta = 0.305$ is the average-pivot contrast (SI Section~\ref{sec:delta}), that is the average pivot ($N = 169{,}737$ applications across 1{,}480 program-years---the 99.1\% of applications with a specialty assignment for which intellectual distance is defined). Points are group means with 95\% confidence intervals; non-pivoters sit $+0.5$ percentage points above their program-year baseline while smaller and larger pivoters sit $0.9$ and $0.7$ points below it. The vertical arrow marks the $\delta = 1.3$ percentage-point gap between non-pivoters and the average pivoter, and the labels mark the modal transitions amongst the bottom and top 10\% of pivoters.}
    \label{fig:empirical_pivot_vs_admission}
\end{figure}

On average, pivoters face lower acceptance rates than non-pivoters. Smaller and larger pivoters faced admission penalties of 1.43 $\pm$ 0.42 pp and 1.17 $\pm$ 0.44 pp, respectively, relative to non-pivoters (differences in mean program-relative admission; $\pm$ denotes the 95\% confidence interval here and throughout). This penalty, which we illustrate in Figure \ref{fig:empirical_pivot_vs_admission}, is estimated relative to other applicants to the same program in the same admissions cycle, which is the comparison an admissions committee actually makes. Next to each group is the most common field transition: smaller pivoters are most often moving from a closely related field (e.g., Electrical \& Electronic Eng. $\rightarrow$ CS: Theory \& Methods), while larger pivoters are more often moving across major disciplinary boundaries (e.g., Physics $\rightarrow$ Astronomy \& Astrophysics). 

Applicants who pivot further are also adversely selected on conventional admissions criteria. Undergraduate GPA declines from non-pivoters ($3.611 \pm 0.003$) to smaller pivoters ($3.553 \pm 0.005$) to larger pivoters ($3.531 \pm 0.005$): relative to non-pivoters, larger pivoters average 0.08 points lower on a 4-point scale, and $d$ is negatively correlated (Spearman $\rho$) with GPA across all applications ($\rho = -0.10$, \textit{p} $< .001$). Undergraduate institution rank is not reliably related to pivot distance ($\rho = +0.004$, \textit{p} = $0.17$). Larger pivoters are, however, more likely to be non-male (47.6\% vs.\ 41.0\% of non-pivoters), more likely to be underrepresented minorities (10.1\% vs.\ 7.6\%), and less likely to be international applicants (47.8\% vs.\ 51.3\%).

\clearpage

\subsection{Pivot penalty in admissions}

Pivoters face a significant penalty in doctoral admissions. In our fully specified linear probability model (Eq. \ref{eq:admissions}; Model 4 of Table \ref{tab:admissions}), the coefficient on the intellectual pivot distance is $-0.026$ (95\% CI: $-0.046$, $-0.006$). Anchored to the average pivot ($\Delta = 0.305$), this corresponds to a 5.3\% relative decline in the probability of admission (0.80 percentage points against the 15.0\% baseline acceptance rate). For context, this is equivalent to a 0.06-point drop in undergraduate GPA (coefficient = 0.141) or losing roughly a third (32.4\%) of the admission premium conferred by attending a Top-100 ranked undergraduate institution. The penalty is robust across a sequence of increasingly stringent specifications: Model 1 is the baseline model; Model 2 adds program-year fixed effects; Model 3 adds core academic achievement variables (undergraduate GPA, GRE-Q and GRE-V percentiles, binned QS World Rank of the undergraduate institution, and an indicator for whether the applicant previously attended University X); and Model 4 adds supplemental academic variables (indicators for whether the applicant possesses a Master's degree and whether they have prior undergraduate research experience through a REU program, and the maximum h-index of the applicant's recommenders) and demographic variables (capturing sex, underrepresented minority status, and US citizenship status). Pivot distance remains negative and significant in every specification.

\begin{table}[!htb]
\centering
\renewcommand\cellalign{t}
\begin{adjustbox}{max width=\textwidth}
\begin{threeparttable}
\begin{tabular}{lcccc}
\toprule
 & \multicolumn{4}{c}{Outcome: Offered Admission} \\
\cmidrule(lr){2-5} 
 & (1) & (2) & (3) & (4) \\
\midrule
\addlinespace
Intellectual-based pivot & \makecell{-0.013** \\ (0.005)} & \makecell{-0.034*** \\ (0.007)} & \makecell{-0.028** \\ (0.009)} & \makecell{-0.026* \\ (0.010)} \\
Undergraduate GPA &  &  & \makecell{0.143*** \\ (0.006)} & \makecell{0.141*** \\ (0.007)} \\
GRE-Q &  &  & \makecell{-0.001*** \\ (0.000)} & \makecell{0.000 \\ (0.000)} \\
GRE-V &  &  & \makecell{0.002*** \\ (0.000)} & \makecell{0.001*** \\ (0.000)} \\
QS World Rank: Top 250 &  &  & \makecell{0.025*** \\ (0.004)} & \makecell{0.020*** \\ (0.004)} \\
QS World Rank: Top 100 &  &  & \makecell{0.028*** \\ (0.004)} & \makecell{0.025*** \\ (0.005)} \\
QS World Rank: Top 50 &  &  & \makecell{0.053*** \\ (0.005)} & \makecell{0.052*** \\ (0.006)} \\
QS World Rank: Top 25 &  &  & \makecell{0.131*** \\ (0.007)} & \makecell{0.117*** \\ (0.007)} \\
University X affiliation &  &  & \makecell{0.019 \\ (0.011)} & \makecell{0.014 \\ (0.013)} \\
Master's degree &  &  &  & \makecell{-0.004 \\ (0.003)} \\
REU/MRE experience &  &  &  & \makecell{0.081*** \\ (0.005)} \\
max(H-index of recommenders) &  &  &  & \makecell{0.001*** \\ (0.000)} \\
Intercept & \makecell{0.150*** \\ (0.001)} &  &  &  \\
\midrule
\addlinespace
Implied effect, average pivot ($\Delta$ = 0.305) & -0.41 pp & -1.05 pp & -0.84 pp & -0.80 pp \\
\midrule
\addlinespace
Program-year fixed effects & - & x & x & x \\
Demographics & - & - & - & x \\
\midrule
\addlinespace
Observations & 169737 & 169643 & 109416 & 80964 \\
S.E. type & iid & by: Program-year & by: Program-year & by: Program-year \\
$R^2$ & 0.000 & 0.058 & 0.114 & 0.137 \\
$R^2$ Within & - & 0.000 & 0.064 & 0.092 \\
\bottomrule
\end{tabular}
\footnotesize Significance levels: $*$ p $<$ 0.05, $**$ p $<$ 0.01, $***$ p $<$ 0.001. Standard errors in parentheses.
\end{threeparttable}
\end{adjustbox}
\caption{\textbf{Admission linear probability models, intellectual pivot distance.} Linear probability models predicting admission (1 = admitted). Model 1 includes only pivot distance. Model 2 adds program-year fixed effects. Model 3 adds academic achievement variables. Model 4 adds supplemental academic variables and demographic controls. Standard errors are clustered by program-year in Models 2--4. The implied effect is the coefficient scaled by $\Delta$, the difference in mean pivot distance between pivoters and non-pivoters (the move from a non-pivoter to the average pivoter).}
\label{tab:admissions}
\end{table}

One possibility is that applicants can offset part of the admissions pivot penalty by securing a recommender in the field of the program to which they apply. Among applicants without a same-field recommender, intellectual pivot distance carries an admission penalty of $-0.035$ per unit (SE $= 0.014$; Table~\ref{tab:rec_lang}). The interaction between pivot distance and having a same-field recommender is positive, $0.026$ (SE $= 0.017$), which would place the conditional penalty at $-0.009$---a 74\% reduction. This interaction is not statistically significant at conventional levels under our primary measure, though it is positive and significant under both alternative pivot measures. Securing a recommender in the target field may therefore be a practical route to lowering the cost of pivoting. This finding parallels prior research in other domains showing that ties with experts in the target field can alleviate pivot penalties \cite{artsParadiseNoveltyLoss2018}. In graduate school admissions, the analogous tie is the recommendation letter writer: a recommender in the target field provides evidence of demonstrated prior engagement with the target discipline---whether through coursework, research assistantships, or co-authorship---which can address evaluators' doubts regarding the applicant's commitment and ability to excel in this new field.

\subsection{Pivot penalty in graduation}

Among admitted applicants who enrolled, greater pivot distance is associated with a substantially lower likelihood of graduating. In our fully specified Cox proportional hazards model (Eq. \ref{eq:graduation}; Model 4 of Table \ref{tab:graduation}), the estimated hazard ratio for a one-unit increase in intellectual pivot distance is 0.690 (95\% CI: 0.493, 0.967). Anchored to the average pivot ($\Delta = 0.305$), this implies a hazard ratio of 0.893---a 10.7\% lower instantaneous rate of graduating at any given time relative to a non-pivoter. This estimate emerges from a sequence of increasingly stringent specifications: Model 1 is the baseline model; Model 2 stratifies by program-year; Model 3 adds academic achievement variables; and Model 4 adds supplemental academic and demographic characteristics. The uncertainty widens as controls are added, but the estimate remains significant throughout.

In contrast to graduation, pivot distance is unrelated to the quantity and citation impact of enrollees' early publications---outcomes important for launching an academic career. Negative binomial models with program-year fixed effects (Table \ref{tab:pubscites_lang}) show no significant association between pivot distance and either the number of post-enrollment publications or the number of citations to those publications. Furthermore, we do not find evidence that pivoting is a ``high risk, high reward'' research strategy. One possibility is that pivoters should exhibit higher variance in outcomes---more frequent rejection, but also more frequent outsized impact \cite{leaheyPerksPerilsInterdisciplinary2018, uzziAtypicalCombinationsScientific2013}---which would manifest as overrepresentation in the upper tail of the outcome distribution even in the absence of a difference in means. We test this with logistic regressions relating enrollees' pivot distance to whether they fall in the top 20\% of publications (Model 1) or citations (Model 2) within their program-year cohort, restricted to program-years with at least five enrollees (SI Section \ref{subsec:upper_tail_performance}). Under our primary measure, pivot distance is not significantly associated with upper-tail performance on either outcome, and neither is the behavioral distance; under the binary indicator the coefficient for upper-tail publications is positive and marginally significant, but it does not replicate under either distance-based measure and no measure shows an association for citations (SI Tables \ref{tab:top20_lang}, \ref{tab:top20_npmi}, \ref{tab:top20_bin}). Taken together with the null average effects above, greater disciplinary distance between prior training and doctoral program does not appear to be a high-reward strategy even for those who are admitted and enroll.

\begin{table}[!htb]
\centering
\begin{adjustbox}{max width=\textwidth}
\renewcommand\cellalign{t}
\begin{threeparttable}
\begin{tabular}{lcccc}
\toprule
 & \multicolumn{4}{c}{Outcome: Graduation (Hazard Ratio)} \\
\cmidrule(lr){2-5}
 & (1) & (2) & (3) & (4) \\
\midrule
\addlinespace
Intellectual-based pivot & \makecell{0.643*** \\ (0.517, 0.798)} & \makecell{0.695** \\ (0.539, 0.897)} & \makecell{0.648* \\ (0.463, 0.908)} & \makecell{0.690* \\ (0.493, 0.967)} \\
Undergraduate GPA &  &  & \makecell{1.078 \\ (0.912, 1.275)} & \makecell{1.166 \\ (0.980, 1.388)} \\
GRE-Q percentile &  &  & \makecell{1.004** \\ (1.001, 1.006)} & \makecell{1.001 \\ (0.998, 1.004)} \\
GRE-V percentile &  &  & \makecell{0.995*** \\ (0.993, 0.997)} & \makecell{0.998 \\ (0.996, 1.001)} \\
QS World Rank: Top 250 &  &  & \makecell{0.993 \\ (0.876, 1.126)} & \makecell{0.983 \\ (0.866, 1.116)} \\
QS World Rank: Top 100 &  &  & \makecell{0.995 \\ (0.877, 1.130)} & \makecell{0.989 \\ (0.870, 1.124)} \\
QS World Rank: Top 50 &  &  & \makecell{0.935 \\ (0.780, 1.121)} & \makecell{0.894 \\ (0.743, 1.077)} \\
QS World Rank: Top 25 &  &  & \makecell{1.141 \\ (0.948, 1.373)} & \makecell{1.166 \\ (0.968, 1.405)} \\
University X affiliation &  &  & \makecell{0.997 \\ (0.769, 1.293)} & \makecell{1.008 \\ (0.777, 1.308)} \\
Master's degree &  &  &  & \makecell{1.194** \\ (1.052, 1.356)} \\
REU/MRE experience &  &  &  & \makecell{0.957 \\ (0.841, 1.090)} \\
max(H-index of recommenders) &  &  &  & \makecell{1.001 \\ (0.999, 1.003)} \\
\midrule
\addlinespace
Implied HR, average pivot ($\Delta$ = 0.305) & 0.874 & 0.895 & 0.876 & 0.893 \\
\midrule
\addlinespace
Program-year & - & x & x & x \\
Demographic characteristics & - & - & - & x \\
\midrule
\addlinespace
Observations & 8731 & 8731 & 4862 & 4862 \\
\bottomrule
\end{tabular}
\footnotesize Significance levels: $*$ p $<$ 0.05, $**$ p $<$ 0.01, $***$ p $<$ 0.001. Format of coefficient cell: Hazard ratio (95\% CI). Models 2--4 stratify by program-year (program\_code + '-' + term), so each program-year has its own baseline hazard. Standard errors are clustered by program-year. The implied HR is the per-unit hazard ratio raised to $\Delta$, the difference in mean pivot distance between pivoters and non-pivoters.
\end{threeparttable}
\end{adjustbox}
\caption{\textbf{Graduation Cox proportional hazards models, intellectual pivot distance.} Cox proportional hazards models predicting graduation (1=graduated). Hazard ratios below 1 indicate lower instantaneous probability of graduating at any given time. Model 1 includes only pivot distance. Model 2 stratifies by program-year (program\_code + '-' + term), allowing each program-year its own baseline hazard. Model 3 adds academic achievement variables. Model 4 adds supplemental academic and demographic controls.}
\label{tab:graduation}
\end{table}

\clearpage

\subsection{Consistency across complementary pivot measures}

We construct two complementary operationalizations of a disciplinary pivot, which confirm the robustness of the pivot penalty in admission and in graduation. In its simplest form, we define the binary pivot indicator, which equals 1 if the applicant's prior degree is in a different Web of Science specialty than the target program and 0 otherwise. This measure is simple and intuitive, but discards information about how far a pivot reaches. We therefore also construct a co-occurrence-based pivot distance measure that captures how often applicants from one field apply to another field. We define this behavioral pivot distance as a rescaling of the normalized pointwise mutual information (NPMI) between the source and target fields, scaled between 0 (most similar) and 1 (most dissimilar), which is a standard measure of association in information theory \cite{bouma2009npmi}.

\begin{figure}[!htb]
    \centering
    \includegraphics[width=\linewidth]{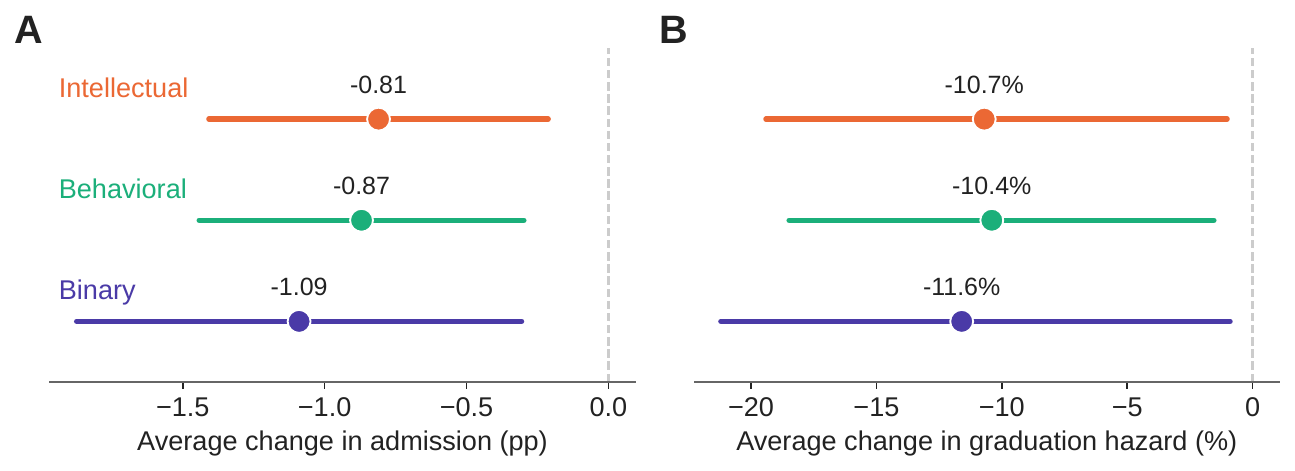}
    \caption{\textbf{Pivot penalties are consistent across all three pivot measures.} Each panel shows the effect of the average pivot---the move from a non-pivoter to the average pivoter---estimated separately under the language-based intellectual pivot distance (orange), the behavioral co-occurrence-based distance (teal), and the binary pivot indicator (violet). Because the three measures are on different scales, each effect is evaluated at that measure's own average-pivot contrast $\Delta$ ($0.305$, $0.265$, and $1.000$, respectively). \textbf{(A)} implied change in admission probability (percentage points), from the fully specified (Model 4) admissions linear probability model. \textbf{(B)} implied change in the graduation hazard (percent), from the fully specified (Model 4) Cox model. Points are estimates with 95\% confidence intervals; the dashed line marks no effect. All six intervals exclude zero.}
    \label{fig:convergence}
\end{figure}

In Figure \ref{fig:convergence} we show that the three measures---the binary pivot indicator, behavioral, and intellectual pivot distance---agree closely on the two central outcomes of admission and graduation. For admission, the average pivot implies a penalty of $-1.09$ percentage points under the binary indicator, $-0.87$ under the behavioral distance, and $-0.80$ under intellectual distance---a range of 0.29 percentage points across three measures constructed on entirely different principles. For graduation, the implied hazard ratios are 0.884, 0.896, and 0.893 respectively, a spread of just over one percentage point. Both penalties are statistically significant under all three measures.

\clearpage

\section{Discussion}
By analyzing applications to the PhD programs at a large R1 university, we contribute four findings to the literature on disciplinary pivots. First, doctoral applicants who apply to a program in a different discipline from their prior area of study (i.e., pivoters) have slightly lower academic achievement on observable dimensions. Second, greater disciplinary distance is associated with lower admission probability: the average pivot corresponds to a 5.3\% lower probability of admission relative to the 15.0\% baseline acceptance rate (0.80 percentage points)---even when controlling for academic achievement and demographic characteristics, and this holds across all three of our pivot measures. Third, having a recommender publishing in the same discipline as the target program appears to mitigate the penalty, cutting the per-unit penalty by roughly three quarters (from $-0.035$ to $-0.009$ under our primary measure, an attenuation that is statistically significant under two of our three measures and consistent in sign under all three), a pattern consistent with a mechanism in which evaluators treat same-field recommenders as evidence of genuine disciplinary engagement. Fourth, conditional on enrollment, applicants who pivot further are significantly less likely to graduate (a 10.7\% lower graduation hazard for the average pivot), again consistently across measures, and we find no reliable evidence that these lower admission and graduation rates are compensated by outsized publication impact on average or in the upper tail. These findings contribute to the emerging body of literature on pivots and research strategies in science by providing evidence for the existence of pivot penalties at the very outset of the academic career pipeline.

This study has a number of limitations that serve as opportunities for future work. First, the data come from a single U.S. university. University X's large size and high research output help it attract many applicants and enrollees but limit generalizability: smaller, less research-intensive universities may treat pivoters more leniently given their less competitive applicant pools, and applicants from smaller institutions may lack access to the breadth of degrees available at University X, making their pivots necessary rather than chosen. Data from other institutions could disentangle the pivot penalty from applicant-pool size and selectivity. Second, our findings on graduation, publications, and citations are conditional on enrollment at University X, introducing selection bias, e.g., collider bias. If admissions committees are effective in selecting students who are more likely to graduate and produce impactful research, then the true pivot penalties for these post-enrollment outcomes are likely more severe than our estimates suggest. Nevertheless, we lack a credible causal identification strategy for post-enrollment outcomes, so we cannot claim that pivoting itself leads to lower likelihood of graduation or publication outcomes. It is plausible that unobserved characteristics (such as a lack of disciplinary identity or a taste for novelty) lead to both pivots and lower program retention. Third, we do not explore the mechanisms to explain why certain applicants (or fields) are more likely to pivot (or contain pivoters) than others. Some applicants may not have access to a range of prior degrees, and some fields may be more welcoming to outsiders than others. Future work could explore the role of field-specific norms and cultures in shaping pivot penalties.
 
Our findings have implications for both future scholars and the scientific research system more broadly. Prospective doctoral applicants considering a pivot face harsher evaluation, but demonstrating substantive engagement with the target discipline and securing recommendation letters from members of the disciplinary community may alleviate evaluators' concerns. For the scientific research system, our results reveal an additional constraint on the flow of new ideas into research communities. While later-career researchers are deterred from pivoting by productivity and citation penalties, aspiring researchers face pivot penalties at the doctoral admissions and graduation stages. Thus, the tension between the community benefits of cognitive diversity and the individual risks of crossing ``out of one's lane'' persists even at the earliest stage of the academic research career.

\clearpage

\section{Materials and Methods}

\subsection{Ethics}
This study was reviewed and approved by the University of Michigan Institutional Review Board (IRB study HUM00225492). Consistent with the IRB's approval of this study as expedited, minimal-risk research on pre-existing records, study-specific informed consent was not required; applicant materials were used under the institution's existing submission acknowledgement along with a supplemental study-specific Memorandum of Understanding (MOU). This MOU specifies the institutional requirements for data security, confidentiality, subject identifiability, and disposal. The acquisition and linkage of third-party data (school rankings, property records, census data, and bibliometric records) to the application records were outlined and approved in the study protocol.

To protect privacy and confidentiality, direct identifiers were used solely for record linkage within the institution's data warehouse and local resources. Direct identifiers were removed before analysis. Data were stored and analyzed exclusively on university-managed resources---the research group's compute cluster, a university-managed desktop workstation with an encrypted local drive, and an approved university Dropbox folder (for initial sharing)---all restricted to IRB-approved study personnel authenticating with secure university credentials. All reported results are aggregate statistics from which no individual applicant or recommender can be identified. Data retention and disposal followed the approved data-management and security plan and the source institution's MOU.

\subsection{Data}

\subsubsection{Doctoral program applications}
We use data from 172,980 applications for the 2013--2023 fall application cycles to 154 different programs (119 fields of study) at a large public R1 university (henceforth ``University X''). The programs span most major disciplines of STEM, social sciences and humanities. Because University X grants a substantial fraction of all doctorates in the U.S. and produces a large fraction of U.S. faculty, it is an ideal site to study the early-career dynamics of disciplinary pivots. 

\subsubsection{Publications}
To study the publication performance of enrollees, we supplement the applications data with bibliometric data from OpenAlex. OpenAlex is an open access research data set containing metadata for over 250M scholarly works, 90M authors, and 100k institutions across all disciplines \cite{priemOpenAlexFullyopenIndex2022}. We match applicants to OpenAlex author records using their name and use University X as the affiliation. 

Our matching procedure is as follows: We first use PyAlex's \texttt{search\_filter} to find OpenAlex author records with display name equal to the enrollee's first and last name. If there are no matches, we assume the enrollee has never published and retain them in our analysis with 0 publications and 0 citations. If there is a single match, we assume that author is the enrollee. If there are multiple matches, we filter the results for those with a present or past affiliation with University X. If this results in no matches, we assume the enrollee has never published; if there is a single match, we assume that author is the enrollee, and if there are still multiple matches, we drop the enrollee from the analysis to reduce error from ambiguous matches. This method results in 64.6\% of enrollees matched, 28.4\% not found, and 7.0\% multiply-matched (and dropped). For each matched author, we obtain from their OpenAlex record (1) the number of works with a publication year on or after their year of enrollment, and (2) the total number of citations to those works. 

\subsubsection{Recommenders}
We analyze applicants' recommenders using Scopus and SciVal. Scopus is a database of 100M+ scholarly works sourced from over 28,000 peer-reviewed titles across a variety of disciplines; SciVal is an analytics tool built on top of Scopus that incorporates additional data on authors, institutions, and countries. We match applicants' recommenders (specifically, recommenders with ``professor'', ``research''/``researcher'', or ``lecturer'' in their title) to the SciVal database via their first name, last name, and affiliation. Each SciVal author record is associated with a collection of subject tags based on the subjects of their published works. We define an author's primary subject as the one in which they have the most publications. 45\% of recommenders were matched to SciVal and have a primary subject area. For analyses involving recommenders, we subset to the 72\% of applicants who have at least one recommender matched to SciVal to avoid our coefficient of interest simply capturing the effect of having a recommender engaged in research.

\subsection{Measures}

\subsubsection{Pivot}
\label{sec:pivot}
We define pivot measures for each application using the subject(s) of the applicant's most recent prior degree (``source subject'') and the subject of the prospective doctoral program (``target subject''), both of which appear in our dataset as text fields and unique code fields. There are 791 distinct prior-degree subject fields and 119 distinct doctoral fields of study.

We map both source subjects and target subjects onto the Web of Science (WoS) Core Collection Subject Categories. In total, there are 254 subject categories, which we call specialties, that match the granularity of our source and target subjects. Each of the 791 prior-degree fields and 119 program fields was assigned to its best-fitting specialty first using a large language model (Claude Opus 4.8). Every assignment was then reviewed by hand, and assignments were revised where the automated label was implausible. Because a single degree may carry two subject areas, a source can map to more than one specialty. Where multiple specialty pairs are possible, we take the minimum distance over all (source specialty $\times$ target specialty) pairs, so an applicant is treated as pivoting only as far as their closest prior training requires. Details of the crosswalk, the review procedure, and the full specialty listing are given in SI Section \ref{sec:crosswalk}.

Our primary pivot measure is the intellectual pivot distance: the language-based distance between the source and target specialties, taken from published pairwise dissimilarities among WoS specialties estimated from the text of roughly $10^{7}$ publication abstracts \cite{diasUsingText2018}. Each subject area is represented by a frequency vector of 20,000 most common non-stop words. A generalized and normalized Jensen-Shannon divergence is computed between two fields using their frequency vectors to yield a dissimilarity measure between 0 and 1. The measure ranges from 0 to 0.843 in our data; non-pivoters---applicants whose prior training and target program share a specialty---are fixed at zero. Transitions between closely related fields (e.g., Biology $\to$ Genetics, Sociology $\to$ Demography) tend to receive small values, whereas transitions between distant fields (e.g., Physics $\to$ History, Chemistry $\to$ Philosophy) tend to receive large values. For example, among pivoting applications, the modal transitions for the bottom 10\%, middle 50\%, and top 10\% of pivot distance are Computer Science, Hardware \& Architecture $\rightarrow$ Computer Science, Theory \& Methods ($N = 987$; $d = 0.07$), Physics, Multidisciplinary $\rightarrow$ Astronomy \& Astrophysics ($N = 1{,}229$; $d = 0.32$), and Psychology $\rightarrow$ Neurosciences ($N = 380$; $d = 0.64$), respectively.

We replicate every analysis using two further measures: a binary pivot indicator equal to 0 when source and target share a specialty and 1 otherwise, and a normalized pointwise mutual information distance computed over the matrix of observed source--target specialty co-occurrences in our own data---a behavioral measure (SI Sections \ref{sec:npmi} and \ref{sec:binary}). The three measures rest on different assumptions and are subject to different biases, yet they converge on the same pivot penalty in admission and in graduation.

Of the 172,980 applications, 1,633 (0.94\%) have a prior-degree field that could not be assigned to a specialty and are excluded from all pivot analyses. A further 1,610 (0.93\%) are successfully mapped but have no published distance for any (source specialty $\times$ target specialty) pair, every such pair involving one of the 29 specialties for which no published distance exists---typically arts and humanities fields. An application whose prior training and target program share a specialty takes $d = 0$ by construction and is retained even when that shared specialty is one of the 29. The intellectual measure is therefore defined for 169,737 applications (98.1\% of applications), and the binary and behavioral measures for 171,347 (99.1\%). Analytic sample sizes fall further in models with academic and demographic controls because of missingness in those covariates. The sample size for each specification is reported in the corresponding table.

\subsubsection{Outcome variables}
Our outcomes of interest are admission, graduation, number of post-enrollment publications, and number of citations to those publications. The first two variables are included in our applications dataset provided by the university and the others are derived from matched OpenAlex records. All outcomes except admission are conditional on enrollment (and therefore also on admission).

\subsubsection{Control variables}
Our models also include other variables that predict admissions but are not of focal interest. These are program-year fixed effects, academic performance variables (undergraduate GPA, GRE score percentiles, QS World Rank of the undergraduate institution), prior affiliation with University X, possession of a Master's degree, prior research experience through a National Science Foundation Research Experience for Undergraduates (NSF-REU) program or similar mentored research experience (MRE), recommender prominence (highest \textit{h}-index among recommenders), and demographic characteristics (sex, U.S. citizenship status, underrepresented minority status). Underrepresented minorities are defined as domestic applicants identifying as Black/African American, Hispanic/Latino, or Native American/Pacific Islander; Asian and Asian-American applicants are not classified as underrepresented minorities under this definition. For publication and citation outcomes, we count only works classified by OpenAlex as journal articles or conference papers; counts are raw (not fractional) across coauthored works.

\subsection{Models}
For simplicity, we predict admission of applicant $i$ applying to program $j$ in year $t$ using the linear probability model
\begin{equation} \label{eq:admissions}
    y_{ijt} = \beta\,\mathrm{pivot}_{ijt}
              + \gamma X_{it} + \alpha_{jt} + \epsilon_{ijt}
\end{equation}
\noindent where $\mathrm{pivot}_{ijt}$ is our intellectual pivot distance (or behavioral or binary pivot), $y_{ijt}$ is an indicator variable (admitted $=1$; rejected $=0$), $X_{it}$ is a vector of covariates with coefficient vector $\gamma$, $\alpha_{jt}$ is a program-year fixed effect, and $\epsilon_{ijt}$ is the error term. Program-year fixed effects subsume all variation in program selectivity and prestige, so no separate program rank control is required. Note also that a pivot is defined relative to the applicant's last degree, so the pivot indicator may change over time as applicants accrue degrees between application cycles. Standard errors are clustered by program-year in all models that include program-year fixed effects or program-year strata.

Because our three pivot measures are defined on different scales, a raw coefficient is not comparable across them, and a one-unit change is in any case larger than any pivot observed in our data. We therefore anchor every effect size to a common contrast, $\Delta$, defined as the difference in mean pivot distance between pivoters and non-pivoters, where pivoter status is given by the binary measure. This is the move from a non-pivoter to the average pivoter. For the binary indicator $\Delta = 1$ by construction; for the behavioral measure, $\Delta = 0.265$; and for intellectual pivot distance, $\Delta = 0.305$. Reported marginal effects are $\beta \times \Delta$, and confidence intervals are scaled by the same factor.

To examine whether having a recommendation letter writer in the same broad field as the program moderates the relationship between pivoting and admissions, we estimate the following linear probability model:
\begin{equation} \label{eq:same_field_rec}
\begin{split}
y_{ijt} =\;& \beta_{1}\,\mathrm{pivot}_{ijt}
+ \beta_{2}\,\mathrm{same\_field\_rec}_{ijt} \\
&+ \beta_{3}\left(\mathrm{pivot}_{ijt}
\times \mathrm{same\_field\_rec}_{ijt}\right)
+ \gamma X_{it} + \alpha_{jt} + \epsilon_{ijt}
\end{split}
\end{equation}
\noindent where $\mathrm{same\_field\_rec}_{ijt}$ equals one if at least one of the applicant's recommenders is in the same broad field as the program and zero otherwise. Since recommender subjects are drawn from the 27 SciVal subjects and programs' subjects are drawn from the 119 doctoral fields of study, we manually map the two sets to determine equivalence (see SI Section \ref{sec:scival_mapping}). We subset our analysis to applicants who have at least one recommender indexed in SciVal to avoid the coefficient of interest capturing the effect of simply having a recommender involved in research. The quantity of interest is $\beta_{1} + \beta_{3}$, interpreted as the pivot penalty among applicants with a target-subject recommender, which we compare to $\beta_1$, the penalty among those without one.

To model graduation (conditional on enrollment), we use a Cox proportional hazards model. A simple linear model is poorly suited for modeling graduation because students who are still enrolled or who leave without graduating do not have an observed completion time. The Cox model retains these right-censored observations by incorporating them into the risk set at each point in time without requiring that their eventual outcome be known. We specify the model as
\begin{equation} \label{eq:graduation}
    \lambda_{ijt}(\tau) = \lambda_{0,jt}(\tau) \exp(\beta \, \mathrm{pivot}_{ijt} + \gamma X_{i})
\end{equation}
\noindent where $\lambda$ is the graduation hazard, $X_{i}$ is the vector of covariates, and $\lambda_{0,jt}(\tau)$ is a program-year-specific baseline hazard. Models 2--4 are stratified by program-year, so each program-year has its own unrestricted baseline hazard and no program-year coefficients are estimated. The coefficient of interest is $\beta$, which, when exponentiated, is interpreted as the hazard ratio for a one-unit increase in pivot distance.

For the number of post-enrollment publications and citations (conditional on enrollment), we fit negative binomial models with the covariates described above, plus program-year fixed effects:
\begin{equation} \label{eq:neg_binom_pub_cites}
    \mu_{ijt} = \exp{(\beta \, \mathrm{pivot}_{ijt} + \gamma X_{i} + \alpha_{jt})}
\end{equation}
\noindent The coefficient of interest is $\beta$. Its exponentiated form, $\exp(\beta)$, is interpreted as the ratio of expected publications (or citations) for a one-unit increase in pivot distance. All statistical tests used to calculate the statistical significance of the parameters reported from these models are two-sided, unless otherwise noted.

\printbibliography

\clearpage

\section*{Acknowledgments}
The authors thank the University X Institutional Research Office for providing the application data.

\subsection*{Funding}
\begin{itemize}
    \item[-] Michigan Institute for Data Science (MIDAS) Propelling Original Data Science (PODS) grant 2023
    \item[-] This material is based upon work supported by the National Science Foundation Graduate Research Fellowship Program under Grant No(s) DGE-2241144. Any opinions, findings, and conclusions or recommendations expressed in this material are those of the author(s) and do not necessarily reflect the views of the National Science Foundation.
\end{itemize}

\subsection*{Author contributions}
Conceptualization: S.X., M.T., D.M.R.; Data curation: S.X., N.D.; Formal analysis: S.X., N.D.; Funding acquisition: M.T., W.S., N.D., D.C.; Investigation: S.X., N.D.; Methodology: S.X., N.D., M.T., D.M.R.; Software: N.D., S.X.; Supervision: M.T., D.M.R., D.C., W.S.; Visualization: N.D.; Writing---original draft: S.X., N.D., M.T.; Writing---review and editing: S.X., N.D., D.C., W.S., D.M.R., M.T.

\subsection*{Competing interests}
The authors declare that they have no competing interests.

\subsection*{Data and materials availability}
The raw PhD-application records analyzed in this study contain personally identifying and protected educational information and cannot be shared publicly, in accordance with the host institution's data-use agreement and the U.S. Family Educational Rights and Privacy Act (FERPA). De-identified aggregate statistics underlying the published figures and tables are available from the corresponding author on reasonable request, subject to host-institution approval. The Web of Science specialty similarities underlying the intellectual pivot distance are from ref. \cite{diasUsingText2018}. Elsevier Scopus and SciVal bibliometric data are commercially available through Elsevier (\url{https://dev.elsevier.com/}). OpenAlex bibliometric data are publicly available from \newline \url{https://openalex.org/}. QS World University Rankings data were obtained from \newline \url{https://www.universityrankings.ch/en/results/QS}.

\clearpage

\setcounter{section}{0}
\setcounter{table}{0}
\setcounter{figure}{0}
\renewcommand{\thetable}{S\arabic{table}}
\renewcommand{\thefigure}{S\arabic{figure}}

\makeatletter
\global\let\maketitle\SI@maketitle
\global\let\@maketitle\SI@@maketitle
\global\let\and\SI@and
\global\let\thanks\SI@thanks
\global\let\@author\SI@author
\global\let\@date\SI@date
\gdef\@title{Supplemental information for\\``The Costs of Early-career Disciplinary Pivots:\\Evidence from PhD Admissions''}
\makeatother

\maketitle

\subsubsection*{This PDF file includes:}
Materials and Methods\\
Figures S1 to S2\\
Tables S1 to S18

\subsubsection*{Other Supplementary Material for this manuscript includes the following:}
Data S1 to S2

\clearpage

\section*{Materials and Methods}
\section{Mapping subjects to Web of Science specialties and SciVal subject areas}
\label{sec:crosswalk}

An applicant's prior-degree subject (the ``source'') and their target doctoral program (the ``target'') appear as short text labels paired with an institutional code in the application records. To allow for comparison across applicants coming from different institutions with different naming conventions, both the source and target degree subjects are mapped to Web of Science (WoS) research specialties. There are 791 unique prior-degree subjects and 119 unique target doctoral programs. This section describes how the mapping was constructed and reviewed, and how a coarser, recommender-facing subject grouping is reconciled with Elsevier's SciVal subject areas for the recommender analysis.

\subsection{The Web of Science subject categories}
\label{sec:wos_hierarchy}

We adopt the Web of Science Core Collection Subject Categories, which contains 254 subject categories, which we call specialties, that match the granularity of our source and target subjects. Language-based pairwise distances (Section~\ref{sec:intellectual}) are published for 225 of the 254 specialties; the remaining 29 specialties have no language coverage (they are retained in full for the behavioral and binary measures). Because the intellectual distance is the minimum over all (source specialty $\times$ target specialty) pairs, an application is dropped from analyses that use it only when no pair has a published distance; pairs involving an uncovered specialty are skipped, and the minimum is taken over those that remain. A transition into the same specialty is a zero-distance transition by construction and is therefore retained at $d = 0$ even when that shared specialty is one of the 29. All specialties are listed in Table~\ref{tab:wos-categories}; the 29 specialties without
language coverage are marked with a dagger ($\dagger$).

\subsection{Mapping prior-degree and program subjects to WoS specialties}
\label{sec:llm_crosswalk}

We built two crosswalks: one from the 791 prior-degree subject fields to WoS specialties, and one from the 119 doctoral fields of study to WoS specialties. Both were constructed in the same way. A large language model (Claude Opus 4.8) was prompted, for each field label, to return the WoS specialty (or specialties; maximum of two) whose scope best matched the field, choosing only from the fixed vocabulary of 254 specialty names; the field's frequency in the data was supplied alongside the label so that high-volume fields could be given closer attention. Because a single degree can carry two subject areas (for example, Electrical \& Computer Engineering maps to both Engineering, Electrical \& Electronic and Computer Science, Hardware \& Architecture), the crosswalk permits up to two specialty assignments per source field; the corresponding rows record both a primary and, where applicable, a secondary specialty. A second specialty is assigned for 68 of the 791 prior-degree fields and 32 of the 119 program fields. The resulting crosswalks draw on 187 distinct specialties on the source side and 88 on the target side, out of the 254 available.

Language models can make errors. Every assignment produced by the language model was therefore reviewed by hand. Author N.D. examined the proposed specialty for each field and revised any assignment whose automated label was implausible or too coarse. All assignments can be seen in the supplemental data, \texttt{prior\_degrees\_WoS\_mapping.csv} and \newline \texttt{applied\_programs\_WoS\_mapping.csv}. Columns denoted by the prefix ``specialty\_number'' contain the specialty numbers used by Dias et al.\ \cite{diasUsingText2018}.

Of the 172,980 total fall matriculating PhD applications, 1,633 (0.94\%) have a prior-degree field with no specialty assignment at all, and a further 1,610 (0.93\%) are successfully mapped but have no published language-based distance for any (source specialty $\times$ target specialty) pair, every such pair involving one of the 29 uncovered specialties---typically arts and humanities fields. Applications that share a specialty with their target program take $d = 0$ by construction and are retained even when that shared specialty is uncovered. Intellectual pivot distance is therefore defined for 169,737 applications (98.1\% of applications), while the behavioral and binary pivot measures, which require only a Web of Science specialty assignment, are defined for 171,347 (99.1\%).

\subsection{Mapping recommender subjects to program fields (SciVal)}
\label{sec:scival_mapping}

The recommender analysis in the main text (the same-field recommender interaction, Eq. \ref{eq:same_field_rec}) requires comparing a recommender's subject of expertise to the applicant's target program. We define the recommender's subject of expertise as the SciVal subject area in which they have published most. SciVal resolves subjects using the All Science Journal Classification (ASJC), consisting of 27 main subject fields. Each of the 119 target PhD program fields is mapped by hand to one of these 27 SciVal subject areas; the full mapping is given in Table~\ref{tab:scival_mapping}. Four SciVal areas (Veterinary, Multidisciplinary, Decision Sciences, and Energy) receive no program, because University X offers no doctoral program in those areas. This coarse SciVal-level matching is used only for the same-field recommender indicator; the pivot measures themselves are computed at the finer WoS-specialty resolution described above.

Since recommenders do not come with a pre-linked publication profile, we first query Elsevier's Scopus/SciVal database using their first name, last name, and affiliation, restricted to recommenders whose title contains ``professor'', ``research''/``researcher'', or ``lecturer''. 514,572 recommendations result in 243,857 unique queries and 123,127 author profiles. 72\% of applicants have at least one recommender indexed in SciVal. An applicant is coded as having a same-field recommender ($\text{same\_field\_rec} = 1$) if the target program's subject area matches the primary subject of at least one of the applicant's recommenders. 

Using these linked bibliometric profiles, we also extract each recommender's \textit{h}-index (if available). Although \textit{h}-index \cite{hirsch2005index} is an imperfect statistic \cite{costas2007hindex, waltman2012inconsistency, adler2009citation, lariviere2013bibliometrics}, it is a quantitative proxy for both visibility and prestige, as well as time in academia. High \textit{h}-index recommenders are more likely to be known by admissions committees which may bring greater influence to their recommendation and their student's odds of admission. Since each applicant has multiple recommenders (typically 3), we take the maximum \textit{h}-index among these as a measure of the applicant's strongest recommendation.

\clearpage

\section{Pivot measures}
\label{sec:measures}

We offer a total of three operationalizations of a disciplinary pivot, each with its own strengths and limitations. The main text reports the results for our primary measure, the intellectual pivot distance, and shows that the two alternative measures---behavioral distance and a binary indicator---yield similar results on the two central outcomes (admission and graduation). This section describes all three measures in detail.

\subsection{Intellectual pivot distance (primary measure)}
\label{sec:intellectual}

Our primary measure is the intellectual pivot distance, a language-based dissimilarity between the source and target specialties. It is taken directly from the published pairwise distances among Web of Science specialties estimated by Dias et al.\ \cite{diasUsingText2018} from the text of roughly $10^{7}$ publication abstracts. It is defined for 169,737 of the 172,980 applications (98.1\%). Missingness is a result of 1,633 applications (0.94\%) having no prior-degree field with a specialty assignment and 1,610 (0.93\%) having no available language measure for any source--target specialty pair. In our data, the language-based distance measure ranges from 0 to 0.843; non-pivoters---applicants whose prior training and target program share a specialty---take the value 0. Transitions between closely related fields (e.g.\ Biology $\to$ Genetics) tend to receive small values, whereas transitions between distant fields (e.g.\ Physics $\to$ History) tend to receive larger values. We prefer this measure because it is derived entirely from the published literature and therefore does not encode applicants' own beliefs about where they are likely to be admitted. Its limitation is that published research is an imperfect proxy for what is taught in undergraduate and master's curricula.

\begin{figure}[!htb]
    \centering
    \includegraphics[width=0.82\linewidth]{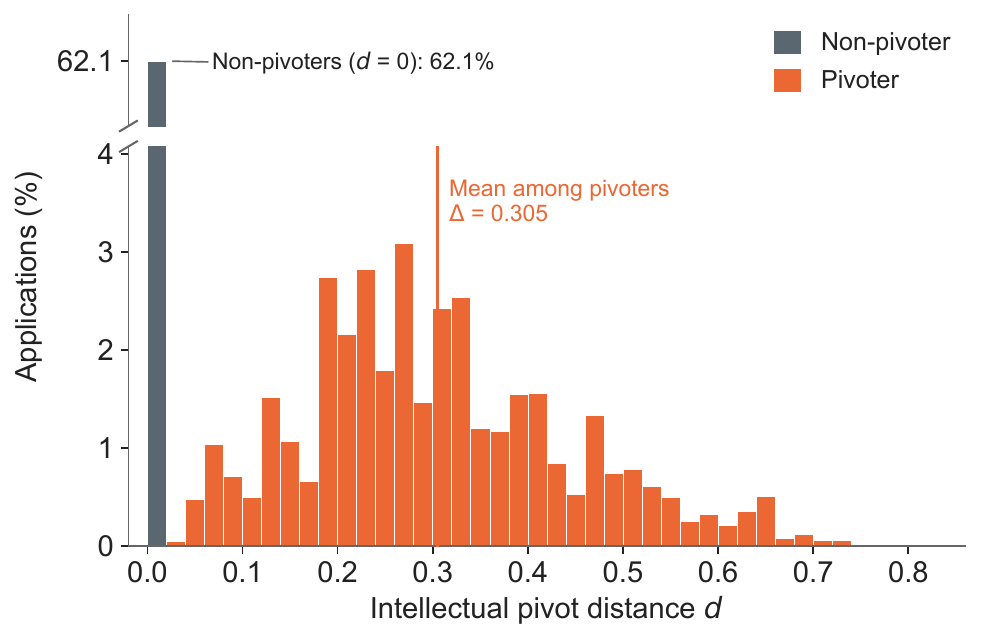}
    \caption{\textbf{Distribution of the intellectual pivot distance.} Binned distribution (bin width $0.02$) of the language-based pivot distance over the $169{,}737$ applications for which it is defined, expressed as a percentage of those applications: non-pivoters (slate; $105{,}402$ applications) and pivoters (orange; $64{,}335$). Non-pivoters share a Web of Science specialty with their target program and therefore take a zero pivot distance by construction, producing the single bin holding $62.1\%$ of applications. The vertical rule marks the average-pivot contrast, $\Delta = 0.305$, between pivoters and non-pivoters (Section~\ref{sec:delta}).}
    \label{fig:dist_intellectual}
\end{figure}

The distribution of the measure across applications is shown in Figure~\ref{fig:dist_intellectual}.

\subsection{Behavioral distance}
\label{sec:npmi}

\begin{figure}[!htb]
    \centering
    \includegraphics[width=0.82\linewidth]{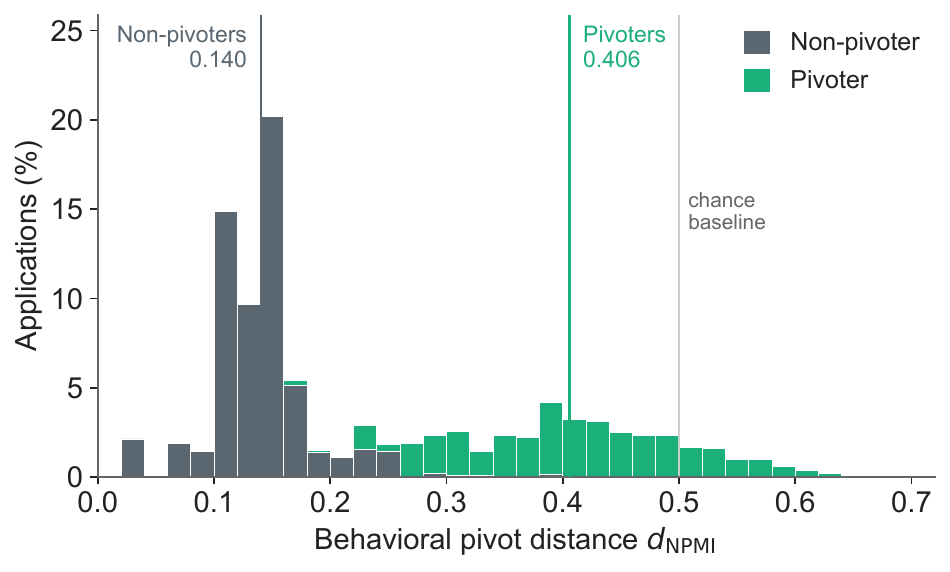}
    \caption{\textbf{Distribution of the behavioral pivot distance.} Binned distribution (bin width $0.02$) of the behavioral-based pivot distance over the $171{,}347$ applications for which it is defined, expressed as a percentage of those applications: non-pivoters (slate; $105{,}402$ applications) and pivoters (green; $65{,}945$). Because the co-occurrence matrix retains its diagonal, same-specialty transitions receive their own empirically estimated distance rather than $0$, so the distribution has no spike at the origin and instead separates into two overlapping modes: non-pivoters concentrate between $0.034$ and $0.456$ (mean $0.140$) and pivoters between $0.166$ and $0.707$ (mean $0.406$). Vertical rules mark the two group means, whose difference is the average-pivot contrast $\Delta = 0.265$ (Section~\ref{sec:delta}). The gray rule at $0.5$ is the chance baseline of the normalized pointwise mutual information distance construction---transitions occurring exactly as often as independence predicts; $6.6\%$ of applications lie beyond it, that is, are rarer than chance.}
    \label{fig:dist_behavioral}
\end{figure}

Our second measure is a behavioral-based one built from applicants' own realized source-to-target transitions rather than from the published literature. Two specialties $X$ and $Y$ are close to the extent that the transition between them occurs more often than chance. Let $n_{XY}$ be the number of applications from source specialty $X$ to target specialty $Y$, with row and column margins $n_{X\cdot}$ and $n_{\cdot Y}$ and grand total $N$. The pointwise mutual information of a transition compares its observed frequency to that expected under independence,
\begin{equation}
    \mathrm{PMI}_{XY} = \ln \frac{n_{XY}\,N}{n_{X\cdot}\,n_{\cdot Y}},
\end{equation}
which we normalize by the self-information of the joint to obtain the normalized pointwise mutual information \cite{bouma2009npmi}
\begin{equation}
    \mathrm{NPMI}_{XY} = \frac{\mathrm{PMI}_{XY}}{-\ln \left(n_{XY}/N\right)} \in [-1, 1],
\end{equation}
and finally map similarity onto a distance,
\begin{equation}
    d_{\mathrm{NPMI},\,XY} = \frac{1 - \mathrm{NPMI}_{XY}}{2}.
\end{equation}
A value of $0$ denotes maximal affinity, $0.5$ the chance baseline, and values approaching $1$ transitions rarer than chance. The measure is defined for the 171,347 applications (99.1\%) with a specialty assignment on both sides, and in our data it ranges from 0.034 to 0.707 with a mean of 0.242. Unlike the original construction, the co-occurrence matrix used here retains the diagonal ($X \to X$ same-specialty flows are kept rather than being forced to zero), so even non-pivoting transitions receive their empirically estimated distance: non-pivoters average 0.140 rather than 0, and pivoters average 0.406. This is a deliberately different convention from the intellectual measure and is one reason the two measures place the non-pivot baseline at slightly different points; the average-pivot contrast $\Delta$ (Section~\ref{sec:delta}) absorbs this difference when effects are compared across measures. The distribution of the behavioral measure is shown in Figure~\ref{fig:dist_behavioral}.

\subsection{Binary pivot indicator}
\label{sec:binary}

Our third measure is the simplest: a binary indicator equal to $0$ when the source and target share a specialty and $1$ otherwise. It discards all information about how far a pivot reaches---treating Biology $\to$ Genetics identically to Physics $\to$ History---but it is free of both the curricular-proxy concern of the intellectual measure and the endogeneity concern of the behavioral measure. It is defined for 171,347 applications (99.1\%) of which 38.5\% are pivots.

\subsection{The average-pivot contrast}
\label{sec:delta}

Because the three measures live on different scales, a raw coefficient is not comparable across them, and a one-unit change is in any case larger than any pivot observed in the data. We therefore anchor every reported effect to a common contrast, $\Delta$, defined as the difference in mean pivot distance between pivoters and non-pivoters (pivoter status being given by the binary measure). This is the effect of moving from a non-pivoter to the average pivoter. By construction $\Delta = 1$ for the binary indicator; $\Delta = 0.265$ for the behavioral measure; and $\Delta = 0.305$ for the intellectual distance. Reported marginal effects are $\beta \times \Delta$ (with confidence intervals scaled by the same factor); for the Cox model the implied hazard ratio is $\exp(\beta)^{\Delta}$.

\clearpage

\section{Supplementary tables for the primary (intellectual) measure}
\label{sec:primary_tables}

The admissions linear probability model (main-text Table~\ref{tab:admissions}) and the graduation Cox model (main-text Table~\ref{tab:graduation}) are tabulated in the main text. Four further tables report, for the primary intellectual pivot distance, the remaining analyses: the same-field recommender interaction (Table~\ref{tab:rec_lang}), a within-applicant comparison of multiple applications to different programs (Table~\ref{tab:multiapp_lang}), the negative binomial models for post-enrollment publications and citations (Table~\ref{tab:pubscites_lang}), and the upper-tail (top-20\%) logistic regressions (Table~\ref{tab:top20_lang}).

\subsection{Same-field recommender interaction}
Estimates of Eq. \ref{eq:same_field_rec} on the 80,963 applications with at least one SciVal-indexed recommender are shown in Table~\ref{tab:rec_lang}. Among applicants without a same-field recommender the per-unit penalty is $-0.035$ (SE $=0.014$); the interaction with having a same-field recommender is $+0.026$ (SE $=0.017$), placing the conditional penalty at $-0.009$, a 74\% attenuation. Anchored to the average pivot ($\Delta = 0.305$), these correspond to $-1.07$ and $-0.27$ percentage points respectively. The interaction is not significant at conventional levels under this measure, but is positive and significant under both alternatives (Tables~\ref{tab:rec_bin} and~\ref{tab:rec_npmi}), suggesting that same-field recommenders may be able to mitigate the pivot penalty.

\subsection{Within-applicant comparison}
Program-year fixed effects compare an applicant to other applicants to the same program in the same cycle, but they do not rule out the possibility that pivoters differ from non-pivoters on unobserved applicant-level traits. We therefore restrict the sample to the applicants who submitted more than one application in the same admissions cycle ($N = 9,575$ applications) and fit a linear mixed model for admission with program-year fixed effects and applicant (student $\times$ term) random effects, so that identification comes substantially from variation in pivot distance within an applicant across the programs they applied to. The coefficient on pivot distance is $-0.019$ (SE $=0.024$), negative and of a magnitude comparable to the pooled estimate but not statistically distinguishable from zero at this much-reduced sample size (Table~\ref{tab:multiapp_lang}). Under the two alternative measures the same specification yields significant negative estimates (Tables~\ref{tab:multiapp_bin} and~\ref{tab:multiapp_npmi}).

\subsection{Publications and citations}
Negative binomial models with program-year fixed effects and standard errors clustered by program-year, fit on enrollees. Pivot distance is not significantly associated with either the post-enrollment publication count ($\beta = 0.256$) or the citation count ($\beta = -0.142$). Full regression coefficients are in Table~\ref{tab:pubscites_lang}.

\subsection{Upper-tail performance}
\label{subsec:upper_tail_performance}
Table~\ref{tab:top20_lang} reports logistic regressions on whether an enrollee falls in the top 20\% of their program-year cohort on publications (Model 1) or citations (Model 2). Cohorts and the top-20\% threshold are computed within program-year, and the sample is restricted to program-years with at least five enrollees so that a top-20\% cell is well defined ($N = 9,005$). Neither coefficient is significant ($0.249$ and $0.119$), giving no evidence that pivoting is a high-variance strategy that pays off in the upper tail.

\clearpage

\section{Replication under the alternative pivot measures}
\label{sec:alt_measures}

To establish that the main findings do not depend on how a pivot is operationalized, we repeated every analysis using the behavioral distance and the binary pivot indicator. Because the three measures live on different scales, every effect below is evaluated at that measure's own average-pivot contrast $\Delta$ (Section~\ref{sec:delta}).

On the two central outcomes the three measures agree closely (main-text Figure \ref{fig:convergence}). For admission, the average pivot implies a penalty of $-1.09$ percentage points under the binary indicator, $-0.87$ under the behavioral distance, and $-0.80$ under intellectual distance---a spread of 0.28 percentage points across three measures built on entirely different principles. For graduation, the implied hazard ratios are $0.884$, $0.896$, and $0.893$ respectively, a spread of just over one percentage point. All six estimates are statistically significant.

The measures differ on three of the smaller-sample secondary analyses, and in each case it is the intellectual measure that is the more conservative. On the within-applicant comparison and the same-field recommender interaction, the two alternative measures yield significant estimates that the intellectual measure does not---a pattern consistent with a real but modest effect estimated more precisely on a co-occurrence-based scale. On the upper-tail (top-20\%) analysis, the binary indicator carries a small positive and marginally significant coefficient for publications ($0.103$, SE $=0.050$; Table~\ref{tab:top20_bin}) where the intellectual and behavioral measures are both null, and no measure shows any association for citations. We do not read the single binary estimate as evidence of an upper-tail advantage: it is one of six upper-tail coefficients, it is significant only at the 5\% level in a model with a pseudo-$R^2$ of essentially zero, and it does not replicate under the two distance-based measures.

\subsection{Behavioral distance}
\label{sec:npmi_tables}

The corresponding replications under the behavioral distance are given in Tables~\ref{tab:admissions_npmi} (admission), \ref{tab:grad_npmi} (graduation), \ref{tab:multiapp_npmi} (within-applicant comparison), \ref{tab:rec_npmi} (same-field recommender interaction), \ref{tab:pubscites_npmi} (publications and citations), and \ref{tab:top20_npmi} (upper-tail performance).

\subsection{Binary pivot indicator}
\label{sec:binary_tables}

The full set of replications under the binary pivot indicator is given in Tables~\ref{tab:admissions_bin} (admission), \ref{tab:grad_bin} (graduation), \ref{tab:multiapp_bin} (within-applicant comparison), \ref{tab:rec_bin} (same-field recommender interaction), \ref{tab:pubscites_bin} (publications and citations), and \ref{tab:top20_bin} (upper-tail performance).

\clearpage

\section*{Figures S1 to S2}

\clearpage

\clearpage

\section*{Tables S1 to S18}


\providecommand{\nold}{\textsuperscript{\dag}}  

\begin{footnotesize}
\setlength{\tabcolsep}{4pt}
\begin{longtable}{@{}>{\raggedright\arraybackslash}p{0.31\textwidth}
                   >{\raggedright\arraybackslash}p{0.31\textwidth}
                   >{\raggedright\arraybackslash}p{0.31\textwidth}@{}}
\caption{\textbf{Web of Science subject categories.} The 254 subject categories of the
Web of Science Core Collection (Science Citation Index Expanded, Social Sciences Citation
Index, and Arts \& Humanities Citation Index), listed alphabetically and read left to right.
Categories marked with a dagger (\dag) have no published language distance and are therefore
excluded from the analyses that require one.}
\label{tab:wos-categories}\\
\toprule
\endfirsthead
\multicolumn{3}{@{}l@{}}{\textit{Table \ref{tab:wos-categories}, continued}}\\
\toprule
\endhead
\midrule
\multicolumn{3}{r@{}}{\textit{Continued on next page}}\\
\endfoot
\bottomrule
\endlastfoot
Acoustics & Agricultural Economics \& Policy & Agricultural Engineering \\
Agriculture, Dairy \& Animal Science & Agriculture, Multidisciplinary & Agronomy \\
Allergy & Anatomy \& Morphology & Andrology \\
Anesthesiology & Anthropology & Archaeology \\
Architecture & Area Studies & Art \\
Asian Studies & Astronomy \& Astrophysics & Audiology \& Speech-Language Pathology\nold \\
Automation \& Control Systems & Behavioral Sciences & Biochemical Research Methods \\
Biochemistry \& Molecular Biology & Biodiversity Conservation & Biology \\
Biophysics & Biotechnology \& Applied Microbiology & Business \\
Business, Finance & Cardiac \& Cardiovascular Systems\nold & Cell Biology \\
Cell \& Tissue Engineering\nold & Chemistry, Analytical & Chemistry, Applied \\
Chemistry, Inorganic \& Nuclear & Chemistry, Medicinal & Chemistry, Multidisciplinary \\
Chemistry, Organic & Chemistry, Physical & Classics \\
Clinical Neurology & Communication & Computer Science, Artificial Intelligence \\
Computer Science, Cybernetics & Computer Science, Hardware \& Architecture & Computer Science, Information Systems \\
Computer Science, Interdisciplinary Applications & Computer Science, Software Engineering & Computer Science, Theory \& Methods \\
Construction \& Building Technology & Criminology \& Penology & Critical Care Medicine \\
Crystallography & Cultural Studies\nold & Dance\nold \\
Demography & Dentistry, Oral Surgery \& Medicine & Dermatology \\
Development Studies\nold & Developmental Biology & Ecology \\
Economics & Education \& Educational Research & Education, Scientific Disciplines \\
Education, Special & Electrochemistry & Emergency Medicine \\
Endocrinology \& Metabolism & Energy \& Fuels & Engineering, Aerospace \\
Engineering, Biomedical & Engineering, Chemical & Engineering, Civil \\
Engineering, Electrical \& Electronic & Engineering, Environmental & Engineering, Geological \\
Engineering, Industrial & Engineering, Manufacturing & Engineering, Marine \\
Engineering, Mechanical & Engineering, Multidisciplinary & Engineering, Ocean \\
Engineering, Petroleum & Entomology & Environmental Sciences \\
Environmental Studies & Ergonomics\nold & Ethics \\
Ethnic Studies\nold & Evolutionary Biology & Family Studies \\
Film, Radio, Television\nold & Fisheries & Folklore \\
Food Science \& Technology & Forestry & Gastroenterology \& Hepatology \\
Genetics \& Heredity & Geochemistry \& Geophysics & Geography \\
Geography, Physical & Geology & Geosciences, Multidisciplinary \\
Geriatrics \& Gerontology & Gerontology & Green \& Sustainable Science \& Technology\nold \\
Health Care Sciences \& Services & Health Policy \& Services & Hematology \\
History & History \& Philosophy of Science & History of Social Sciences \\
Horticulture & Hospitality, Leisure, Sport \& Tourism & Humanities, Multidisciplinary \\
Imaging Science \& Photographic Technology & Immunology & Industrial Relations \& Labor \\
Infectious Diseases & Information Science \& Library Science & Instruments \& Instrumentation \\
Integrative \& Complementary Medicine & International Relations & Language \& Linguistics\nold \\
Law & Limnology & Linguistics \\
Literary Reviews\nold & Literary Theory \& Criticism\nold & Literature \\
Literature, African, Australian, Canadian\nold & Literature, American\nold & Literature, British Isles\nold \\
Literature, German, Dutch, Scandinavian & Literature, Romance & Literature, Slavic\nold \\
Logic\nold & Management & Marine \& Freshwater Biology \\
Materials Science, Biomaterials & Materials Science, Ceramics & Materials Science, Characterization \& Testing\nold \\
Materials Science, Coatings \& Films & Materials Science, Composites & Materials Science, Multidisciplinary \\
Materials Science, Paper \& Wood & Materials Science, Textiles & Mathematical \& Computational Biology \\
Mathematics & Mathematics, Applied & Mathematics, Interdisciplinary Applications \\
Mechanics & Medical Ethics\nold & Medical Informatics \\
Medical Laboratory Technology & Medicine, General \& Internal & Medicine, Legal \\
Medicine, Research \& Experimental & Medieval \& Renaissance Studies & Metallurgy \& Metallurgical Engineering \\
Meteorology \& Atmospheric Sciences & Microbiology & Microscopy \\
Mineralogy & Mining \& Mineral Processing & Multidisciplinary Sciences \\
Music & Mycology & Nanoscience \& Nanotechnology \\
Neuroimaging\nold & Neurosciences & Nuclear Science \& Technology \\
Nursing & Nutrition \& Dietetics & Obstetrics \& Gynecology \\
Oceanography & Oncology & Operations Research \& Management Science \\
Ophthalmology & Optics & Ornithology \\
Orthopedics & Otorhinolaryngology & Paleontology \\
Parasitology\nold & Pathology & Pediatrics \\
Peripheral Vascular Disease\nold & Pharmacology \& Pharmacy & Philosophy \\
Physics, Applied & Physics, Atomic, Molecular \& Chemical & Physics, Condensed Matter \\
Physics, Fluids \& Plasmas & Physics, Mathematical & Physics, Multidisciplinary \\
Physics, Nuclear & Physics, Particles \& Fields & Physiology \\
Plant Sciences & Poetry\nold & Political Science \\
Polymer Science & Primary Health Care\nold & Psychiatry \\
Psychology & Psychology, Applied & Psychology, Biological \\
Psychology, Clinical & Psychology, Developmental & Psychology, Educational \\
Psychology, Experimental & Psychology, Mathematical\nold & Psychology, Multidisciplinary \\
Psychology, Psychoanalysis & Psychology, Social & Public Administration \\
Public, Environmental \& Occupational Health & Quantum Science \& Technology\nold & Radiology, Nuclear Medicine \& Medical Imaging \\
Regional \& Urban Planning & Rehabilitation & Religion \\
Remote Sensing & Reproductive Biology & Respiratory System \\
Rheumatology & Robotics & Social Issues \\
Social Sciences, Biomedical & Social Sciences, Interdisciplinary & Social Sciences, Mathematical Methods \\
Social Work & Sociology & Soil Science \\
Spectroscopy & Sport Sciences & Statistics \& Probability \\
Substance Abuse & Surgery & Telecommunications \\
Theater & Thermodynamics & Toxicology \\
Transplantation & Transportation\nold & Transportation Science \& Technology \\
Tropical Medicine & Urban Studies & Urology \& Nephrology \\
Veterinary Sciences & Virology & Water Resources \\
Women's Studies\nold & Zoology &  \\
\end{longtable}
\end{footnotesize}

\clearpage
\begin{footnotesize}
\setlength{\tabcolsep}{4pt}
\begin{longtable}{@{}p{0.30\textwidth} p{0.64\textwidth}@{}}
\caption{\textbf{PhD program fields of study mapped to SciVal subject areas.} Each of the 119 doctoral fields of study is assigned by hand to one of the 27 SciVal (Scopus ASJC top-level) subject areas. A recommender's primary subject is the SciVal area in which they have published most; an applicant has a same-field recommender if the target program's SciVal area matches the primary subject of at least one of their recommenders. Four SciVal areas receive no program because University X offers no doctoral program in them.}\\
\label{tab:scival_mapping}\\
\toprule
\textbf{SciVal subject area} & \textbf{PhD fields of study} \\
\midrule
\endfirsthead
\multicolumn{2}{l}{\emph{(Table \ref{tab:scival_mapping}, continued)}}\\[2pt]
\toprule
\textbf{SciVal subject area} & \textbf{PhD fields of study} \\
\midrule
\endhead
\midrule \multicolumn{2}{r}{\emph{continued on next page}}\\
\endfoot
\bottomrule
\endlastfoot
Agricultural \& Biological Sciences (AGRI) & Nutritional Sciences \\
\addlinespace
Arts \& Humanities (ARTS) & Architecture; Asian Languages \& Cultures; Comparative Literature; Design Science; English Language \& Literature; Germanic Languages \& Literatures; History of Art; Landscape Architecture; Linguistics \& Romanic Languages \& Literatures; Music: Composition; Music: Composition \& Music Theory; Music: Conducting; Music: Jazz \& Contemporary Improvisation; Music: Music Education; Music: Musicology; Music: Performance; Music: Performing Arts Technology; Music: Theory; Philosophy; Romance Languages \& Literatures: French; Romance Languages \& Literatures: Italian; Romance Languages \& Literatures: Spanish; Slavic Languages \& Literatures \\
\addlinespace
Biochemistry, Genetics \& Molecular Biology (BIOC) & Biological Chemistry; Biophysics; Cancer Biology; Cell \& Developmental Biology; Cellular \& Molecular Biology; Chemical Biology; Genetics \& Genomics; Molec., Cell. \& Develop. Bio.; Molecular \& Cellular Pathology; Molecular \& Integrative Physiology \\
\addlinespace
Business, Management \& Accounting (BUSI) & Business \& Economics; Business Administration \\
\addlinespace
Chemical Engineering (CENG) & Chemical Engin. \\
\addlinespace
Chemistry (CHEM) & Chemistry \\
\addlinespace
Computer Science (COMP) & Computer Science \& Engin. \\
\addlinespace
Dentistry (DENT) & Oral Health Sciences \\
\addlinespace
Earth \& Planetary Sciences (EART) & Climate \& Space Sci. \& Engin.; Earth \& Environmental Sciences \\
\addlinespace
Economics, Econometrics \& Finance (ECON) & Economics \\
\addlinespace
Engineering (ENGI) & Aerospace Engin.; Civil Engin.; Electrical \& Computer Engineering; Engineering Education Research; Industrial \& Operations Engin.; Mechanical Engin.; Naval Arch. \& Marine Engin.; Nuclear Engin. \& Radiol. Sci.; Nuclear Science; Robotics \\
\addlinespace
Environmental Science (ENVI) & Ecology \& Evolutionary Biology; Environment \& Sustainability; Environmental Engin. \\
\addlinespace
Health Professions (HEAL) & Environmental Health Sciences; Epidemiological Science; Health Behavior \& Health Education; Health Infrastructures \& Learning Sys.; Health Services Organization \& Policy; Kinesiology; Movement Science; Sport Management \\
\addlinespace
Immunology \& Microbiology (IMMU) & Immunology; Microbiology \& Immunology \\
\addlinespace
Materials Science (MATE) & Macromolecular Science \& Engin.; Materials Science \& Engin. \\
\addlinespace
Mathematics (MATH) & Applied \& Interdisciplinary Mathematics; Mathematics; Statistics \\
\addlinespace
Medicine (MEDI) & Bioinformatics; Biomedical Engin.; Biostatistics; Medical Scientist Training Program; Program In Biomedical Sciences; Toxicology \\
\addlinespace
Neuroscience (NEUR) & Neuroscience \\
\addlinespace
Nursing (NURS) & Nursing \\
\addlinespace
Pharmacology, Toxicology \& Pharmaceutics (PHAR) & Clinical Pharmacy Translational Sciences; Medicinal Chemistry; Pharmaceutical Sciences; Pharmacology \\
\addlinespace
Physics \& Astronomy (PHYS) & Applied Physics; Astronomy \& Astrophysics; Physics; Space \& Planetary Physics \\
\addlinespace
Psychology (PSYC) & Psychology \\
\addlinespace
Social Sciences (SOCI) & American Culture; Ancient History; Anthropology; Anthropology \& History; Classical Art \& Archaeology; Classical Studies; Communication \& Media; Education \& Psychology; Educational Studies; English \& Education; English \& Women's \& Gender Studies; Film,Television, \& Media; Higher Education; History; History \& Women's \& Gender Studies; Information; Linguistics; Middle East Studies; Political Science; Psychology \& Women's \& Gender Studies; Public Policy \& Economics; Public Policy \& Political Science; Public Policy \& Sociology; Social Work \& Anthropology; Social Work \& Economics; Social Work \& Political Science; Social Work \& Psychology; Social Work \& Sociology; Sociology; Survey \& Data Science; Urban \& Regional Planning \\
\addlinespace
Veterinary (VETE) & \emph{(none assigned)} \\
\addlinespace
Multidisciplinary (MULT) & \emph{(none assigned)} \\
\addlinespace
Decision Sciences (DECI) & \emph{(none assigned)} \\
\addlinespace
Energy (ENER) & \emph{(none assigned)} \\
\end{longtable}
\end{footnotesize}

\clearpage

\begin{table}[!htb] \centering
\begin{adjustbox}{max width=\textwidth}
\begin{threeparttable}
\begin{tabular}{@{\extracolsep{5pt}}lc}
\\[-1.8ex]\hline
\\[-1.8ex] Outcome: & Offered Admission \\
\\[-1.8ex] & (1) \\
\hline \\[-1.8ex]
 Intellectual-based pivot & -0.019$^{}$ \\
& (0.024) \\ [0.2cm]
 Intercept & -0.010$^{}$ \\
& (0.213) \\ [0.2cm]
\midrule
\addlinespace
Program-year fixed effects & x \\
Applicant random effects & x \\
\hline \\[-1.8ex]
 Observations & 9575 \\
 Residual Std. Error & 0.314 (df=8671) \\
\hline
\end{tabular}
\footnotesize Significance levels: $*$ p $<$ 0.05, $**$ p $<$ 0.01, $***$ p $<$ 0.001. Standard errors in parentheses.
\end{threeparttable}
\end{adjustbox}
\caption{\textbf{Within-applicant admission comparison, intellectual pivot distance.} Linear mixed model predicting admission (1 = admitted) among applicants who applied to multiple doctoral programs in the same admissions cycle, with program-year (program $\times$ term) fixed effects and applicant random effects.}
\label{tab:multiapp_lang}
\end{table}

\clearpage
\begin{table}[!htb] \centering
\begin{threeparttable}
\begin{tabular}{lc}
\toprule
 Outcome: & Offered Admission \\
\midrule
\addlinespace
Intellectual-based pivot & \makecell{-0.035* \\ (0.014)} \\
Same-field recommender & \makecell{0.011** \\ (0.004)} \\
Intellectual-based pivot $\times$ Same-field recommender & \makecell{0.026 \\ (0.017)} \\
Undergraduate GPA & \makecell{0.140*** \\ (0.007)} \\
GRE-Q percentile & \makecell{0.000 \\ (0.000)} \\
GRE-V percentile & \makecell{0.001*** \\ (0.000)} \\
QS World Rank: Top 250 & \makecell{0.020*** \\ (0.004)} \\
QS World Rank: Top 100 & \makecell{0.025*** \\ (0.005)} \\
QS World Rank: Top 50 & \makecell{0.052*** \\ (0.006)} \\
QS World Rank: Top 25 & \makecell{0.117*** \\ (0.007)} \\
University X affiliation & \makecell{0.014 \\ (0.013)} \\
Master's degree & \makecell{-0.005 \\ (0.003)} \\
REU/MRE experience & \makecell{0.081*** \\ (0.005)} \\
max(H-index of recommenders) & \makecell{0.001*** \\ (0.000)} \\
\midrule
\addlinespace
Program-year fixed effects & x \\
Demographics & x \\
\midrule
\addlinespace
Observations & 80963 \\
S.E. type & by: Program-year \\
$R^2$ & 0.138 \\
$R^2$ Within & 0.092 \\
\bottomrule
\end{tabular}
\footnotesize Significance levels: $*$ p $<$ 0.05, $**$ p $<$ 0.01, $***$ p $<$ 0.001. Standard errors in parentheses.
\end{threeparttable}
\caption{\textbf{Same-field recommender interaction, intellectual pivot distance.} Linear probability model estimating the moderating effect on admission of having a recommender in the target discipline. The variable same-field recommender equals 1 if at least one of the applicant's recommenders has a primary Scopus$/$SciVal subject in the same broad field as the target program. The sample is restricted to applicants with at least one recommender indexed in SciVal. Standard errors are clustered by program-year.}
\label{tab:rec_lang}
\end{table}

\clearpage
\begin{table}[!htb]
\centering
\begin{adjustbox}{max width=\textwidth}
\begingroup
\centering
\begin{threeparttable}
   \begin{tabular}{lcc}
      \toprule
      Dependent variables:         & \# Publications  & \# Citations \\    
      Model:                       & (1)              & (2)\\  
      \midrule 
      Intellectual-based pivot     & 0.2562           & -0.1418\\   
                                   & (0.1651)         & (0.2291)\\   
      Undergraduate GPA            & 0.1223           & 0.1811\\   
                                   & (0.0793)         & (0.1343)\\   
      GRE-Q                        & 0.0028           & 0.0050\\   
                                   & (0.0017)         & (0.0026)\\   
      GRE-V                        & -0.0014          & -0.0030\\   
                                   & (0.0014)         & (0.0023)\\   
      QS World Rank: Top 250       & 0.1036           & 0.2126\\   
                                   & (0.0737)         & (0.1127)\\   
      QS World Rank: Top 100       & 0.0435           & 0.1539\\   
                                   & (0.0833)         & (0.1363)\\   
      QS World Rank: Top 50        & 0.3501$^{**}$    & 0.5129$^{**}$\\   
                                   & (0.1216)         & (0.1836)\\   
      QS World Rank: Top 25        & 0.1802           & 0.2238\\   
                                   & (0.1039)         & (0.1802)\\   
      University X affiliation     & -0.0209          & 0.0165\\   
                                   & (0.1289)         & (0.1989)\\   
      Master's degree              & 0.1206           & 0.1607\\   
                                   & (0.0670)         & (0.1060)\\   
      REU/MRE experience           & 0.0013           & 0.0466\\   
                                   & (0.0634)         & (0.0974)\\   
      max(H-index of recommenders) & 0.0015           & 0.0041$^{*}$\\   
                                   & (0.0012)         & (0.0018)\\   
      \midrule 
      Program-year fixed effects   & x                & x\\  
      Demographics                 & x                & x\\  
      \midrule 
      S.E. type                    & by: Program-year & by: Program-year\\  
      Observations                 & 4,604            & 4,523\\  
      Squared Correlation          & 0.19049          & 0.15337\\  
      Pseudo R$^2$                 & 0.05058          & 0.03648\\  
      BIC                          & 34,506.5         & 48,236.6\\  
      Over-dispersion              & 0.60348          & 0.27019\\  
      \bottomrule
   \end{tabular}
   \footnotesize Significance levels: $*$ p $<$ 0.05, $**$ p $<$ 0.01, $***$ p $<$ 0.001. Standard errors in parentheses.
\end{threeparttable}
\par\endgroup
\end{adjustbox}
\caption{\textbf{Post-enrollment publications and citations, intellectual pivot distance.} Negative binomial models predicting the number of post-enrollment publications (Model 1) and citations to those publications (Model 2), using the language-based intellectual pivot distance. Coefficients are on the log scale; exponentiated coefficients give incidence-rate ratios. Standard errors are clustered by program-year.}
\label{tab:pubscites_lang}
\end{table}

\clearpage
\begin{table}[!htb]
\centering
\renewcommand\cellalign{t}
\begin{adjustbox}{max width=\textwidth}
\begin{threeparttable}
\begin{tabular}{lcc}
\toprule
Dependent variables: & Top 20\% publications & Top 20\% citations \\
\cmidrule(lr){2-3} 
Model: & (1) & (2) \\
\midrule
\addlinespace
Intellectual-based pivot & \makecell{0.249 \\ (0.143)} & \makecell{0.119 \\ (0.145)} \\
Intercept & \makecell{-1.630*** \\ (0.020)} & \makecell{-1.579*** \\ (0.020)} \\
\midrule
\addlinespace
Observations & 9005 & 9005 \\
S.E. type & by: Program-year & by: Program-year \\
Pseudo $R^2$ & 0.000 & 0.000 \\
\bottomrule
\end{tabular}
\footnotesize Significance levels: $*$ p $<$ 0.05, $**$ p $<$ 0.01, $***$ p $<$ 0.001. Standard errors in parentheses.
\end{threeparttable}
\end{adjustbox}
\caption{\textbf{Upper-tail (top 20\%) publications and citations, intellectual pivot distance.} Logistic regressions predicting whether an enrollee's publications (Model 1) or citations (Model 2) fall in the top 20\% of their program-year cohort, restricted to program-years with at least five enrollees. Coefficients are on the log-odds scale.}
\label{tab:top20_lang}
\end{table}

\clearpage

\begin{table}[!htb]
\centering
\renewcommand\cellalign{t}
\begin{adjustbox}{max width=\textwidth}
\begin{threeparttable}
\begin{tabular}{lcccc}
\toprule
 & \multicolumn{4}{c}{Outcome: Offered Admission} \\
\cmidrule(lr){2-5} 
 & (1) & (2) & (3) & (4) \\
\midrule
\addlinespace
Behavioral-based pivot & \makecell{-0.026*** \\ (0.006)} & \makecell{-0.062*** \\ (0.009)} & \makecell{-0.039*** \\ (0.010)} & \makecell{-0.033** \\ (0.011)} \\
Undergraduate GPA &  &  & \makecell{0.142*** \\ (0.006)} & \makecell{0.140*** \\ (0.007)} \\
GRE-Q &  &  & \makecell{-0.001*** \\ (0.000)} & \makecell{0.000 \\ (0.000)} \\
GRE-V &  &  & \makecell{0.002*** \\ (0.000)} & \makecell{0.001*** \\ (0.000)} \\
QS World Rank: Top 250 &  &  & \makecell{0.025*** \\ (0.004)} & \makecell{0.020*** \\ (0.004)} \\
QS World Rank: Top 100 &  &  & \makecell{0.029*** \\ (0.004)} & \makecell{0.026*** \\ (0.005)} \\
QS World Rank: Top 50 &  &  & \makecell{0.053*** \\ (0.005)} & \makecell{0.052*** \\ (0.006)} \\
QS World Rank: Top 25 &  &  & \makecell{0.132*** \\ (0.007)} & \makecell{0.118*** \\ (0.007)} \\
University X affiliation &  &  & \makecell{0.015 \\ (0.011)} & \makecell{0.011 \\ (0.013)} \\
Master's degree &  &  &  & \makecell{-0.004 \\ (0.003)} \\
REU/MRE experience &  &  &  & \makecell{0.081*** \\ (0.005)} \\
max(H-index of recommenders) &  &  &  & \makecell{0.001*** \\ (0.000)} \\
Intercept & \makecell{0.155*** \\ (0.002)} &  &  &  \\
\midrule
\addlinespace
Implied effect, average pivot ($\Delta$ = 0.265) & -0.69 pp & -1.65 pp & -1.04 pp & -0.87 pp \\
\midrule
\addlinespace
Program-year fixed effects & - & x & x & x \\
Demographics & - & - & - & x \\
\midrule
\addlinespace
Observations & 171347 & 171252 & 110246 & 81559 \\
S.E. type & iid & by: Program-year & by: Program-year & by: Program-year \\
$R^2$ & 0.000 & 0.058 & 0.114 & 0.137 \\
$R^2$ Within & - & 0.001 & 0.063 & 0.091 \\
\bottomrule
\end{tabular}
\footnotesize Significance levels: $*$ p $<$ 0.05, $**$ p $<$ 0.01, $***$ p $<$ 0.001. Standard errors in parentheses.
\end{threeparttable}
\end{adjustbox}
\caption{\textbf{Admission linear probability models, behavioral pivot distance.} Linear probability models predicting admission (1 = admitted). Model 1 includes only pivot distance. Model 2 adds program-year fixed effects. Model 3 adds academic achievement variables. Model 4 adds supplemental academic variables and demographic controls. Standard errors are clustered by program-year in Models 2--4. The implied effect is the coefficient scaled by $\Delta$, the difference in mean pivot distance between pivoters and non-pivoters (the move from a non-pivoter to the average pivoter).}
\label{tab:admissions_npmi}
\end{table}

\clearpage
\begin{table}[!htb]
\centering
\begin{adjustbox}{max width=\textwidth}
\renewcommand\cellalign{t}
\begin{threeparttable}
\begin{tabular}{lcccc}
\toprule
 & \multicolumn{4}{c}{Outcome: Graduation (Hazard Ratio)} \\
\cmidrule(lr){2-5}
 & (1) & (2) & (3) & (4) \\
\midrule
\addlinespace
Behavioral-based pivot & \makecell{0.407*** \\ (0.311, 0.533)} & \makecell{0.634** \\ (0.479, 0.837)} & \makecell{0.647* \\ (0.450, 0.931)} & \makecell{0.661* \\ (0.463, 0.943)} \\
Undergraduate GPA &  &  & \makecell{1.072 \\ (0.907, 1.266)} & \makecell{1.162 \\ (0.977, 1.382)} \\
GRE-Q percentile &  &  & \makecell{1.004** \\ (1.001, 1.006)} & \makecell{1.001 \\ (0.998, 1.004)} \\
GRE-V percentile &  &  & \makecell{0.995*** \\ (0.993, 0.997)} & \makecell{0.998 \\ (0.996, 1.001)} \\
QS World Rank: Top 250 &  &  & \makecell{0.990 \\ (0.874, 1.122)} & \makecell{0.980 \\ (0.863, 1.112)} \\
QS World Rank: Top 100 &  &  & \makecell{0.990 \\ (0.873, 1.123)} & \makecell{0.984 \\ (0.867, 1.117)} \\
QS World Rank: Top 50 &  &  & \makecell{0.938 \\ (0.782, 1.125)} & \makecell{0.897 \\ (0.745, 1.080)} \\
QS World Rank: Top 25 &  &  & \makecell{1.132 \\ (0.941, 1.361)} & \makecell{1.160 \\ (0.963, 1.396)} \\
University X affiliation &  &  & \makecell{1.003 \\ (0.774, 1.300)} & \makecell{1.016 \\ (0.783, 1.318)} \\
Master's degree &  &  &  & \makecell{1.196** \\ (1.055, 1.356)} \\
REU/MRE experience &  &  &  & \makecell{0.953 \\ (0.836, 1.086)} \\
max(H-index of recommenders) &  &  &  & \makecell{1.001 \\ (0.999, 1.003)} \\
\midrule
\addlinespace
Implied HR, average pivot ($\Delta$ = 0.265) & 0.788 & 0.886 & 0.891 & 0.896 \\
\midrule
\addlinespace
Program-year & - & x & x & x \\
Demographic characteristics & - & - & - & x \\
\midrule
\addlinespace
Observations & 8813 & 8813 & 4892 & 4892 \\
\bottomrule
\end{tabular}
\footnotesize Significance levels: $*$ p $<$ 0.05, $**$ p $<$ 0.01, $***$ p $<$ 0.001. Format of coefficient cell: Hazard ratio (95\% CI). Models 2--4 stratify by program-year (program\_code + '-' + term), so each program-year has its own baseline hazard. Standard errors are clustered by program-year. The implied HR is the per-unit hazard ratio raised to $\Delta$, the difference in mean pivot distance between pivoters and non-pivoters.
\end{threeparttable}
\end{adjustbox}
\caption{\textbf{Graduation Cox proportional hazards models, behavioral pivot distance.} Cox proportional hazards models predicting graduation (1=graduated). Hazard ratios below 1 indicate lower instantaneous probability of graduating at any given time. Model 1 includes only pivot distance. Model 2 stratifies by program-year (program\_code + '-' + term), allowing each program-year its own baseline hazard. Model 3 adds academic achievement variables. Model 4 adds supplemental academic and demographic controls.}
\label{tab:grad_npmi}
\end{table}

\clearpage
\begin{table}[!htb] \centering
\begin{adjustbox}{max width=\textwidth}
\begin{threeparttable}
\begin{tabular}{@{\extracolsep{5pt}}lc}
\\[-1.8ex]\hline
\\[-1.8ex] Outcome: & Offered Admission \\
\\[-1.8ex] & (1) \\
\hline \\[-1.8ex]
 Behavioral-based pivot & -0.072$^{**}$ \\
& (0.028) \\ [0.2cm]
 Intercept & 0.012$^{}$ \\
& (0.212) \\ [0.2cm]
\midrule
\addlinespace
Program-year fixed effects & x \\
Applicant random effects & x \\
\hline \\[-1.8ex]
 Observations & 9854 \\
 Residual Std. Error & 0.312 (df=8921) \\
\hline
\end{tabular}
\footnotesize Significance levels: $*$ p $<$ 0.05, $**$ p $<$ 0.01, $***$ p $<$ 0.001. Standard errors in parentheses.
\end{threeparttable}
\end{adjustbox}
\caption{\textbf{Within-applicant admission comparison, behavioral pivot distance.} Linear mixed model predicting admission (1 = admitted) among applicants who applied to multiple doctoral programs in the same admissions cycle, with program-year (program $\times$ term) fixed effects and applicant random effects.}
\label{tab:multiapp_npmi}
\end{table}

\clearpage
\begin{table}[!htb] \centering
\begin{threeparttable}
\begin{tabular}{lcc}
\toprule
 Outcome: & Offered Admission \\
\midrule
\addlinespace
Behavioral-based pivot & \makecell{-0.052** \\ (0.016)} \\
Same-field recommender & \makecell{0.003 \\ (0.006)} \\
Behavioral-based pivot $\times$ Same-field recommender & \makecell{0.047* \\ (0.019)} \\
Undergraduate GPA & \makecell{0.139*** \\ (0.007)} \\
GRE-Q percentile & \makecell{0.000 \\ (0.000)} \\
GRE-V percentile & \makecell{0.001*** \\ (0.000)} \\
QS World Rank: Top 250 & \makecell{0.020*** \\ (0.004)} \\
QS World Rank: Top 100 & \makecell{0.026*** \\ (0.005)} \\
QS World Rank: Top 50 & \makecell{0.052*** \\ (0.006)} \\
QS World Rank: Top 25 & \makecell{0.118*** \\ (0.007)} \\
University X affiliation & \makecell{0.011 \\ (0.013)} \\
Master's degree & \makecell{-0.004 \\ (0.003)} \\
REU/MRE experience & \makecell{0.081*** \\ (0.005)} \\
max(H-index of recommenders) & \makecell{0.001*** \\ (0.000)} \\
\midrule
\addlinespace
Program-year fixed effects & x \\
Demographics & x \\
\midrule
\addlinespace
Observations & 81558 \\
S.E. type & by: Program-year \\
$R^2$ & 0.137 \\
$R^2$ Within & 0.091 \\
\bottomrule
\end{tabular}
\footnotesize Significance levels: $*$ p $<$ 0.05, $**$ p $<$ 0.01, $***$ p $<$ 0.001. Standard errors in parentheses.
\end{threeparttable}
\caption{\textbf{Same-field recommender interaction, behavioral pivot distance.} Linear probability model estimating the moderating effect on admission of having a recommender in the target discipline. The variable same-field recommender equals 1 if at least one of the applicant's recommenders has a primary Scopus$/$SciVal subject in the same broad field as the target program. The sample is restricted to applicants with at least one recommender indexed in SciVal. Standard errors are clustered by program-year.}
\label{tab:rec_npmi}
\end{table}

\clearpage
\begin{table}[!htb]
\centering
\begin{adjustbox}{max width=\textwidth}
\begingroup
\centering
\begin{threeparttable}
   \begin{tabular}{lcc}
      \toprule
      Dependent variables:         & \# Publications  & \# Citations \\    
      Model:                       & (1)              & (2)\\  
      \midrule 
      Behavioral-based pivot       & 0.1594           & -0.1282\\   
                                   & (0.1783)         & (0.2854)\\   
      Undergraduate GPA            & 0.1126           & 0.1694\\   
                                   & (0.0800)         & (0.1348)\\   
      GRE-Q                        & 0.0028           & 0.0051\\   
                                   & (0.0017)         & (0.0026)\\   
      GRE-V                        & -0.0013          & -0.0028\\   
                                   & (0.0014)         & (0.0023)\\   
      QS World Rank: Top 250       & 0.0977           & 0.2074\\   
                                   & (0.0737)         & (0.1132)\\   
      QS World Rank: Top 100       & 0.0434           & 0.1625\\   
                                   & (0.0836)         & (0.1364)\\   
      QS World Rank: Top 50        & 0.3503$^{**}$    & 0.5144$^{**}$\\   
                                   & (0.1213)         & (0.1844)\\   
      QS World Rank: Top 25        & 0.1638           & 0.2002\\   
                                   & (0.1041)         & (0.1808)\\   
      University X affiliation     & 0.0046           & 0.0584\\   
                                   & (0.1306)         & (0.2025)\\   
      Master's degree              & 0.1124           & 0.1604\\   
                                   & (0.0670)         & (0.1057)\\   
      REU/MRE experience           & 0.0061           & 0.0443\\   
                                   & (0.0633)         & (0.0973)\\   
      max(H-index of recommenders) & 0.0015           & 0.0042$^{*}$\\   
                                   & (0.0012)         & (0.0018)\\   
      \midrule 
      Program-year fixed effects   & x                & x\\  
      Demographics                 & x                & x\\  
      \midrule 
      S.E. type                    & by: Program-year & by: Program-year\\  
      Observations                 & 4,639            & 4,550\\  
      Squared Correlation          & 0.19524          & 0.14930\\  
      Pseudo R$^2$                 & 0.05155          & 0.03692\\  
      BIC                          & 34,700.1         & 48,414.6\\  
      Over-dispersion              & 0.60363          & 0.26974\\  
      \bottomrule
   \end{tabular}
   \footnotesize Significance levels: $*$ p $<$ 0.05, $**$ p $<$ 0.01, $***$ p $<$ 0.001. Standard errors in parentheses.
\end{threeparttable}
\par\endgroup
\end{adjustbox}
\caption{\textbf{Post-enrollment publications and citations, behavioral pivot distance.} Negative binomial models predicting the number of post-enrollment publications (Model 1) and citations to those publications (Model 2), using the behavioral distance $d_{\mathrm{NPMI}}$. Coefficients are on the log scale; exponentiated coefficients give incidence-rate ratios. Standard errors are clustered by program-year.}
\label{tab:pubscites_npmi}
\end{table}

\clearpage
\begin{table}[!htb]
\centering
\renewcommand\cellalign{t}
\begin{adjustbox}{max width=\textwidth}
\begin{threeparttable}
\begin{tabular}{lcc}
\toprule
Dependent variables: & Top 20\% publications & Top 20\% citations \\
\cmidrule(lr){2-3} 
Model: & (1) & (2) \\
\midrule
\addlinespace
Behavioral-based pivot & \makecell{0.214 \\ (0.164)} & \makecell{0.270 \\ (0.171)} \\
Intercept & \makecell{-1.657*** \\ (0.043)} & \makecell{-1.638*** \\ (0.044)} \\
\midrule
\addlinespace
Observations & 9095 & 9095 \\
S.E. type & by: Program-year & by: Program-year \\
Pseudo $R^2$ & 0.000 & 0.000 \\
\bottomrule
\end{tabular}
\footnotesize Significance levels: $*$ p $<$ 0.05, $**$ p $<$ 0.01, $***$ p $<$ 0.001. Standard errors in parentheses.
\end{threeparttable}
\end{adjustbox}
\caption{\textbf{Upper-tail (top 20\%) publications and citations, behavioral pivot distance.} Logistic regressions predicting whether an enrollee's publications (Model 1) or citations (Model 2) fall in the top 20\% of their program-year cohort, restricted to program-years with at least five enrollees. Coefficients are on the log-odds scale.}
\label{tab:top20_npmi}
\end{table}

\clearpage

\begin{table}[!htb]
\centering
\renewcommand\cellalign{t}
\begin{adjustbox}{max width=\textwidth}
\begin{threeparttable}
\begin{tabular}{lcccc}
\toprule
 & \multicolumn{4}{c}{Outcome: Offered Admission} \\
\cmidrule(lr){2-5} 
 & (1) & (2) & (3) & (4) \\
\midrule
\addlinespace
Binary pivot & \makecell{-0.003 \\ (0.002)} & \makecell{-0.018*** \\ (0.003)} & \makecell{-0.013*** \\ (0.003)} & \makecell{-0.011** \\ (0.004)} \\
Undergraduate GPA &  &  & \makecell{0.142*** \\ (0.006)} & \makecell{0.140*** \\ (0.007)} \\
GRE-Q &  &  & \makecell{-0.001*** \\ (0.000)} & \makecell{0.000 \\ (0.000)} \\
GRE-V &  &  & \makecell{0.002*** \\ (0.000)} & \makecell{0.001*** \\ (0.000)} \\
QS World Rank: Top 250 &  &  & \makecell{0.025*** \\ (0.004)} & \makecell{0.020*** \\ (0.004)} \\
QS World Rank: Top 100 &  &  & \makecell{0.029*** \\ (0.004)} & \makecell{0.025*** \\ (0.005)} \\
QS World Rank: Top 50 &  &  & \makecell{0.053*** \\ (0.005)} & \makecell{0.052*** \\ (0.006)} \\
QS World Rank: Top 25 &  &  & \makecell{0.132*** \\ (0.007)} & \makecell{0.118*** \\ (0.007)} \\
University X affiliation &  &  & \makecell{0.015 \\ (0.011)} & \makecell{0.011 \\ (0.013)} \\
Master's degree &  &  &  & \makecell{-0.004 \\ (0.003)} \\
REU/MRE experience &  &  &  & \makecell{0.081*** \\ (0.005)} \\
max(H-index of recommenders) &  &  &  & \makecell{0.001*** \\ (0.000)} \\
Intercept & \makecell{0.150*** \\ (0.001)} &  &  &  \\
\midrule
\addlinespace
Implied effect, average pivot ($\Delta$ = 1.000) & -0.26 pp & -1.76 pp & -1.28 pp & -1.09 pp \\
\midrule
\addlinespace
Program-year fixed effects & - & x & x & x \\
Demographics & - & - & - & x \\
\midrule
\addlinespace
Observations & 171347 & 171252 & 110246 & 81559 \\
S.E. type & iid & by: Program-year & by: Program-year & by: Program-year \\
$R^2$ & 0.000 & 0.058 & 0.114 & 0.137 \\
$R^2$ Within & - & 0.000 & 0.063 & 0.091 \\
\bottomrule
\end{tabular}
\footnotesize Significance levels: $*$ p $<$ 0.05, $**$ p $<$ 0.01, $***$ p $<$ 0.001. Standard errors in parentheses.
\end{threeparttable}
\end{adjustbox}
\caption{\textbf{Admission linear probability models, binary pivot indicator.} Linear probability models predicting admission (1 = admitted). Pivot indicator equals 1 if the applicant pivoted specialty (else 0). Model 1 includes only pivot distance. Model 2 adds program-year fixed effects. Model 3 adds academic achievement variables. Model 4 adds supplemental academic variables and demographic controls. Standard errors are clustered by program-year in Models 2--4. The implied effect is the coefficient scaled by $\Delta$, the difference in mean pivot distance between pivoters and non-pivoters (the move from a non-pivoter to the average pivoter).}
\label{tab:admissions_bin}
\end{table}

\clearpage
\begin{table}[!htb]
\centering
\begin{adjustbox}{max width=\textwidth}
\renewcommand\cellalign{t}
\begin{threeparttable}
\begin{tabular}{lcccc}
\toprule
 & \multicolumn{4}{c}{Outcome: Graduation (Hazard Ratio)} \\
\cmidrule(lr){2-5}
 & (1) & (2) & (3) & (4) \\
\midrule
\addlinespace
Binary pivot & \makecell{0.836*** \\ (0.777, 0.900)} & \makecell{0.878** \\ (0.804, 0.960)} & \makecell{0.872* \\ (0.776, 0.981)} & \makecell{0.884* \\ (0.788, 0.991)} \\
Undergraduate GPA &  &  & \makecell{1.073 \\ (0.909, 1.267)} & \makecell{1.163 \\ (0.978, 1.381)} \\
GRE-Q percentile &  &  & \makecell{1.004** \\ (1.001, 1.006)} & \makecell{1.001 \\ (0.998, 1.004)} \\
GRE-V percentile &  &  & \makecell{0.995*** \\ (0.993, 0.997)} & \makecell{0.998 \\ (0.996, 1.001)} \\
QS World Rank: Top 250 &  &  & \makecell{0.988 \\ (0.871, 1.120)} & \makecell{0.978 \\ (0.861, 1.110)} \\
QS World Rank: Top 100 &  &  & \makecell{0.991 \\ (0.873, 1.125)} & \makecell{0.984 \\ (0.866, 1.118)} \\
QS World Rank: Top 50 &  &  & \makecell{0.938 \\ (0.783, 1.124)} & \makecell{0.897 \\ (0.745, 1.079)} \\
QS World Rank: Top 25 &  &  & \makecell{1.130 \\ (0.939, 1.361)} & \makecell{1.157 \\ (0.960, 1.394)} \\
University X affiliation &  &  & \makecell{1.003 \\ (0.774, 1.300)} & \makecell{1.016 \\ (0.784, 1.318)} \\
Master's degree &  &  &  & \makecell{1.191** \\ (1.051, 1.350)} \\
REU/MRE experience &  &  &  & \makecell{0.952 \\ (0.836, 1.084)} \\
max(H-index of recommenders) &  &  &  & \makecell{1.001 \\ (0.999, 1.003)} \\
\midrule
\addlinespace
Implied HR, average pivot ($\Delta$ = 1.000) & 0.836 & 0.878 & 0.872 & 0.884 \\
\midrule
\addlinespace
Program-year & - & x & x & x \\
Demographic characteristics & - & - & - & x \\
\midrule
\addlinespace
Observations & 8813 & 8813 & 4892 & 4892 \\
\bottomrule
\end{tabular}
\footnotesize Significance levels: $*$ p $<$ 0.05, $**$ p $<$ 0.01, $***$ p $<$ 0.001. Format of coefficient cell: Hazard ratio (95\% CI). Models 2--4 stratify by program-year (program\_code + '-' + term), so each program-year has its own baseline hazard. Standard errors are clustered by program-year. The implied HR is the per-unit hazard ratio raised to $\Delta$, the difference in mean pivot distance between pivoters and non-pivoters.
\end{threeparttable}
\end{adjustbox}
\caption{\textbf{Graduation Cox proportional hazards models, binary pivot indicator.} Cox proportional hazards models predicting graduation (1=graduated). Hazard ratios below 1 indicate lower instantaneous probability of graduating at any given time. Model 1 includes only pivot distance. Model 2 stratifies by program-year (program\_code + '-' + term), allowing each program-year its own baseline hazard. Model 3 adds academic achievement variables. Model 4 adds supplemental academic and demographic controls.}
\label{tab:grad_bin}
\end{table}

\clearpage
\begin{table}[!htb] \centering
\begin{adjustbox}{max width=\textwidth}
\begin{threeparttable}
\begin{tabular}{@{\extracolsep{5pt}}lc}
\\[-1.8ex]\hline
\\[-1.8ex] Outcome: & Offered Admission \\
\\[-1.8ex] & (1) \\
\hline \\[-1.8ex]
 Binary pivot & -0.022$^{*}$ \\
& (0.009) \\ [0.2cm]
 Intercept & -0.003$^{}$ \\
& (0.212) \\ [0.2cm]
\midrule
\addlinespace
Program-year fixed effects & x \\
Applicant random effects & x \\
\hline \\[-1.8ex]
 Observations & 9854 \\
 Residual Std. Error & 0.312 (df=8921) \\
\hline
\end{tabular}
\footnotesize Significance levels: $*$ p $<$ 0.05, $**$ p $<$ 0.01, $***$ p $<$ 0.001. Standard errors in parentheses.
\end{threeparttable}
\end{adjustbox}
\caption{\textbf{Within-applicant admission comparison, binary pivot indicator.} Linear mixed model predicting admission (1 = admitted) among applicants who applied to multiple doctoral programs in the same admissions cycle, with program-year (program $\times$ term) fixed effects and applicant random effects.}
\label{tab:multiapp_bin}
\end{table}

\clearpage
\begin{table}[!htb] \centering
\begin{threeparttable}
\begin{tabular}{lcc}
\toprule
 Outcome: & Offered Admission \\
\midrule
\addlinespace
Binary pivot & \makecell{-0.017*** \\ (0.005)} \\
Same-field recommender & \makecell{0.009* \\ (0.004)} \\
Binary pivot $\times$ Same-field recommender & \makecell{0.013* \\ (0.006)} \\
Undergraduate GPA & \makecell{0.139*** \\ (0.007)} \\
GRE-Q percentile & \makecell{0.000 \\ (0.000)} \\
GRE-V percentile & \makecell{0.001*** \\ (0.000)} \\
QS World Rank: Top 250 & \makecell{0.020*** \\ (0.004)} \\
QS World Rank: Top 100 & \makecell{0.026*** \\ (0.005)} \\
QS World Rank: Top 50 & \makecell{0.052*** \\ (0.006)} \\
QS World Rank: Top 25 & \makecell{0.118*** \\ (0.007)} \\
University X affiliation & \makecell{0.011 \\ (0.013)} \\
Master's degree & \makecell{-0.004 \\ (0.003)} \\
REU/MRE experience & \makecell{0.081*** \\ (0.005)} \\
max(H-index of recommenders) & \makecell{0.001*** \\ (0.000)} \\
\midrule
\addlinespace
Program-year fixed effects & x \\
Demographics & x \\
\midrule
\addlinespace
Observations & 81558 \\
S.E. type & by: Program-year \\
$R^2$ & 0.137 \\
$R^2$ Within & 0.091 \\
\bottomrule
\end{tabular}
\footnotesize Significance levels: $*$ p $<$ 0.05, $**$ p $<$ 0.01, $***$ p $<$ 0.001. Standard errors in parentheses.
\end{threeparttable}
\caption{\textbf{Same-field recommender interaction, binary pivot indicator.} Linear probability model estimating the moderating effect on admission of having a recommender in the target discipline. The variable same-field recommender equals 1 if at least one of the applicant's recommenders has a primary Scopus$/$SciVal subject in the same broad field as the target program. The sample is restricted to applicants with at least one recommender indexed in SciVal. Standard errors are clustered by program-year.}
\label{tab:rec_bin}
\end{table}

\clearpage
\begin{table}[!htb]
\centering
\begin{adjustbox}{max width=\textwidth}
\begingroup
\centering
\begin{threeparttable}
   \begin{tabular}{lcc}
      \toprule
      Dependent variables:         & \# Publications  & \# Citations \\    
      Model:                       & (1)              & (2)\\  
      \midrule 
      Binary pivot                 & 0.0877           & -0.0178\\   
                                   & (0.0552)         & (0.0859)\\   
      Undergraduate GPA            & 0.1130           & 0.1713\\   
                                   & (0.0796)         & (0.1347)\\   
      GRE-Q                        & 0.0029           & 0.0051$^{*}$\\   
                                   & (0.0017)         & (0.0026)\\   
      GRE-V                        & -0.0013          & -0.0028\\   
                                   & (0.0014)         & (0.0023)\\   
      QS World Rank: Top 250       & 0.1013           & 0.2084\\   
                                   & (0.0737)         & (0.1132)\\   
      QS World Rank: Top 100       & 0.0465           & 0.1631\\   
                                   & (0.0836)         & (0.1364)\\   
      QS World Rank: Top 50        & 0.3500$^{**}$    & 0.5168$^{**}$\\   
                                   & (0.1210)         & (0.1844)\\   
      QS World Rank: Top 25        & 0.1651           & 0.2018\\   
                                   & (0.1039)         & (0.1810)\\   
      University X affiliation     & 0.0072           & 0.0562\\   
                                   & (0.1306)         & (0.2020)\\   
      Master's degree              & 0.1167           & 0.1611\\   
                                   & (0.0670)         & (0.1061)\\   
      REU/MRE experience           & 0.0044           & 0.0432\\   
                                   & (0.0633)         & (0.0974)\\   
      max(H-index of recommenders) & 0.0015           & 0.0042$^{*}$\\   
                                   & (0.0012)         & (0.0018)\\   
      \midrule 
      Program-year fixed effects   & x                & x\\  
      Demographics                 & x                & x\\  
      \midrule 
      S.E. type                    & by: Program-year & by: Program-year\\  
      Observations                 & 4,639            & 4,550\\  
      Squared Correlation          & 0.19391          & 0.14860\\  
      Pseudo R$^2$                 & 0.05161          & 0.03691\\  
      BIC                          & 34,698.4         & 48,414.8\\  
      Over-dispersion              & 0.60394          & 0.26973\\  
      \bottomrule
   \end{tabular}
   \footnotesize Significance levels: $*$ p $<$ 0.05, $**$ p $<$ 0.01, $***$ p $<$ 0.001. Standard errors in parentheses.
\end{threeparttable}
\par\endgroup
\end{adjustbox}
\caption{\textbf{Post-enrollment publications and citations, binary pivot indicator.} Negative binomial models predicting the number of post-enrollment publications (Model 1) and citations to those publications (Model 2), using the binary pivot indicator. Coefficients are on the log scale; exponentiated coefficients give incidence-rate ratios. Standard errors are clustered by program-year.}
\label{tab:pubscites_bin}
\end{table}

\clearpage
\begin{table}[!htb]
\centering
\renewcommand\cellalign{t}
\begin{adjustbox}{max width=\textwidth}
\begin{threeparttable}
\begin{tabular}{lcc}
\toprule
Dependent variables: & Top 20\% publications & Top 20\% citations \\
\cmidrule(lr){2-3} 
Model: & (1) & (2) \\
\midrule
\addlinespace
Binary pivot & \makecell{0.103* \\ (0.050)} & \makecell{0.068 \\ (0.051)} \\
Intercept & \makecell{-1.646*** \\ (0.023)} & \makecell{-1.599*** \\ (0.023)} \\
\midrule
\addlinespace
Observations & 9095 & 9095 \\
S.E. type & by: Program-year & by: Program-year \\
Pseudo $R^2$ & 0.000 & 0.000 \\
\bottomrule
\end{tabular}
\footnotesize Significance levels: $*$ p $<$ 0.05, $**$ p $<$ 0.01, $***$ p $<$ 0.001. Standard errors in parentheses.
\end{threeparttable}
\end{adjustbox}
\caption{\textbf{Upper-tail (top 20\%) publications and citations, binary pivot indicator.} Logistic regressions predicting whether an enrollee's publications (Model 1) or citations (Model 2) fall in the top 20\% of their program-year cohort, restricted to program-years with at least five enrollees. Coefficients are on the log-odds scale.}
\label{tab:top20_bin}
\end{table}

\clearpage

\section*{Data S1 to S2}
\textbf{Data S1: Prior-degree program mapping to Web of Science subject areas.} Each of the 791 unique prior degree names are mapped (or not mapped) to Web of Science subject areas. In some cases, a prior degree is mapped to a secondary subject area (subject\_area\_2). Columns denoted by the prefix ``specialty\_number'' contain the specialty numbers used by Dias et al.\ \cite{diasUsingText2018}.

\vspace{1em}

\noindent\textbf{Data S2: Applied PhD program mapping to Web of Science subject areas.} Each of the 119 unique PhD programs fields of study were mapped (or not mapped) to Web of Science subject areas. In some cases, a prior degree is mapped to a secondary subject area (subject\_area\_2). Columns denoted by the prefix ``specialty\_number'' contain the specialty numbers used by Dias et al.\ \cite{diasUsingText2018}.

\end{document}